\PassOptionsToPackage{table}{xcolor}
\documentclass[twocolumn,a4paper]{aastex631}
\usepackage[paperwidth=21cm,paperheight=29.7cm,margin=2.0cm]{geometry}
 \let\tabular\savetabular
 \usepackage{amsmath}
 \usepackage{amssymb}
 \usepackage{natbib}
 \usepackage{multirow}
 \usepackage{graphicx}\graphicspath{{graphics/}}
 \usepackage{color}
 \usepackage{epstopdf}
\received{\today}
\revised{\today}
\accepted{\today}
\submitjournal{ApJ}

\shorttitle{Evolution of Protoclusters}
\shortauthors{Remus, Dolag, \& Dannerbauer}

\begin{document}

\title{The Young and the Wild: What happens to Protoclusters forming at \boldmath$z\approx4$?}

\correspondingauthor{Rhea-Silvia Remus}
\email{rhea@usm.lmu.de}
\author{Rhea-Silvia Remus}
\affil{Universit\"ats-Sternwarte M\"unchen, Fakult\"at f\"ur Physik, Ludwig-Maximilians-Universit\"at, Scheinerstr.~1, D-81679 M\"unchen, Germany}
\author{Klaus Dolag}
\affil{Universit\"ats-Sternwarte M\"unchen, Fakult\"at f\"ur Physik, Ludwig-Maximilians-Universit\"at, Scheinerstr.~1, D-81679 M\"unchen, Germany}
\affil{Max-Planck-Institute for Astrophysics, Karl-Schwarzschild-Str.~1, D-85748 Garching, Germany}
\author{Helmut Dannerbauer}
\affil{Instituto de Astrof\'isica de Canarias, E-38205 La Laguna, Tenerife, Spain}
\affil{Universidad de La Laguna Dpto. Astrof\'isica, E-38206 La Laguna, Tenerife, Spain}

\begin{abstract}
Using one of the largest volumes of the hydrodynamical cosmological
simulation suit {\it Magneticum}, we study the evolution of protoclusters identified
at redshift $\approx4$, with properties similar to {\it SPT2349-56}. We identify 42 protoclusters in the simulation, as massive and equally rich in substructures as observed, confirming that these structures are already virialized. The dynamics of the internally fast rotating member galaxies within these protoclusters resembles observations, merging rapidly to form the cores of the BCGs of the assembling clusters. Half of the gas reservoir of these structures is in a hot phase, with the metal-enrichment at a very early stage. These systems show a good agreement with the observed amount of cold star-forming gas, largely enriched to solar values. We predict that some of the member galaxies are already quenched at $z\approx4$, rendering them undetectable through measurements of their gas reservoir.
Tracing the evolution of protoclusters reveals that none of the typical mass indicators at high redshift are good tracers to predict the present-day mass of the system. We find that none of the simulated protoclusters with properties as {\it SPT2349-56} at $z=4.3$, are among the top ten most massive clusters at redshift $z=0$, with some barely reaching masses of $M\approx2\times10^{14}M_\odot$. Although the average star-formation and mass-growth rates in the simulated galaxies match observations at high redshift reasonably well, the simulation fails to reproduce the extremely high total star-formation rates within observed protoclusters, indicating that the sub-grid models are lacking the ability to reproduce higher star-formation efficiency (or lower depletion timescales).
\end{abstract}


\section{Introduction}
Overdensities of galaxies at very high redshifts have been observed in increasingly large amounts in the last few years, reaching redshifts as high as $z=6$ and more. Assuming that those massive agglomerations of galaxies are the cores of structures that will collapse into very massive galaxy clusters at present day, these structures have been named {\it protoclusters} \citep[see][for an overview]{overzier:2016}.
Structures that will eventually collapes into a massive galaxy cluster by $z=0$ are stretched out over several tenth to hundreds of Mpc \cite[e.g.]{muldrew:2015}, and the galaxy overdensities that are observed thus are only the (possibly already collapsed) cores of these structures. However, naming has been imprecise here, with the term protocluster being used usually for what are the cores of the assembling structures.

Some of these observed protocluster cores reach masses high enough to challenge predictions from $\Lambda$CDM cosmological simulations, for example the two massive protoclusters observed at $z\approx4$, namely {\it SPT2349-56} at $z=4.3$ \citep{miller:2018,rotermund:2021} with a total mass of more than $1\times10^{13}M_\odot$, and the even more massive protocluster reported by \citet{oteo:2018} at $z=4.0$ with a total mass above $4\times10^{13}M_\odot$. Both of these protocluster cores have large numbers of member galaxies with extremely high total star formation rates of more than $6000 M_\odot/yr$.
Even more challenging is the structure reported by \citet{chanchaiworawit:2019} at $z=6.5$ with a virialized core mass of $M_\mathrm{200} \approx4.06\times10^{13}M_\odot$. 
Many more such overdensities at redshifts of $z=4$ and higher have been recently reported \citep[e.g.,][]{ouchi:2005,toshikawa:2012,toshikawa:2014,harikane:2019,calvi:2019,toshikawa:2020,calvi:2021}, with still large masses but not as extreme.

At lower redshift of about $z \approx 2$, several protocluster cores have been already observed. One of the first reported and by now best studied protoclusters at this redshift is the so-called Spiderweb-galaxy at $z=2.16$, which actually consists of several galaxies with extremely high star formation rates \citep{dannerbauer:2014,shimakawa:2014,shimakawa:2018}, larger than what is reported for the star forming main sequence at these redshifts \citep{santini:2017,pearson:2018}, and often they are associated with massive star forming submillimeter galaxies \citep[e.g.,][]{zhang:2022}.
Extremely high star formation rates for both the whole protocluster core but also the individual galaxies in these cores have been confirmed for other protocluster cores as well from redshifts $4<z<2$ \citep[e.g.,][]{umehata:2015,wang:2016,kubo:2016,kubo:2017,wang:2018}, especially through observations of CO with ALMA.
\citet{strazzullo:2018b} especially used CO observations of clusters at $z\approx 2$ to show that the star formation rates in these clusters are strongly enhanced compared to the field.
However, while \citet{aoyama:2022} reported for their protocluster at $z\approx2.5$ rather high star formation rates, they cannot confirm that these are higher than what is seen for star forming galaxies of the same mass in the field.

While many of the galaxies in such protoclusters seem to have enhanced star formation rates, some already quenched galaxies have also been reported. For example, \citep{kubo:2013} already reported a quiescent fraction of about $20\dots 50\%$ in a protocluster region at $z=3.1$, while \citet{mcconachie:2022} confirm even a quiescent fraction of $70\%$ in a protocluster core at $z=3.37$, and \citet{shi:2019} found an enhancement of quiescent galaxies in a protocluster structure at $z=3.78$.
A clear environmental dependence of the quenched fraction of galaxies can be already found around $3<z<2$ \citep{kodama:2007,yonekura:2021}, and for redshifts between $1<z<2$, quenched fractions have been shown to increase with lower redshifts in clusters \citep[e.g.,][]{cooke:2019,sarron:2021} especially compared to the field \citep{cooke:2019}. The red sequence buildup is clearly apparent in galaxy clusters $z=2$ and below \citep{strazzullo:2013,hatch:2017,strazzullo:2016,ando:2022}, suggesting that agglomeration of red sequence galaxies could be good tracers for (proto)clusters at high redshifts \citep{strazzullo:2015}, albeit detections of such quiescent galaxies are still rare above $z=3$, see for example \citet{kubo:2021}.
This is in agreement with simulations that report the morphology density relation to be building up around $z=2$ \citep{teklu:2017}.

From the simulation side, the evolution of galaxy clusters has been predicted to high redshifts from models and dark matter only cosmological simulations so far \citep[e.g.,][]{chiang:2013,muldrew:2015}, as extremely large fully hydrodynamical cosmological simulations are expensive and thus still rare, but required to reproduce massive collapsed protoclusters at high redshifts. However, zoom-simulations of individual galaxy clusters have been used to study the star formation properties of protocluster galaxies \citep{bassini:2020}, but also especially the build-up of the cores of todays most massive galaxies, the brightest cluster galaxies (BCGs) \citep{ragone:2018,rennehan:2020}.

There are various and sometimes not concordantly used denotations of protoclusters and protocluster cores in the literature, with the exact terminology still a matter of debate. As we are using data from simulations in this work, we have the full 3D and evolution information for all our simulated structures, and thus we will stick to a strictly physically motivated definition, calling the virialized\footnote{Note that in simulations virialized structures are defined via a density contrast predicted from spherical Top-Hat models. The protocluster structures, however, are fast growing, so the virial ratio here should also including the surface term and should not be confused with being a static system. Nevertheless, the velocity dispersion of the member galaxies typically reflects the virial velocity of the halo in good approximation.} regions identified in the simulations at a given redshift {\it protocluster}, as they are already bound structures with a common dark matter halo. We will call the Lagrangian region, which comprised everything that will end up in the final galaxy cluster at $z=0$, {\it protocluster region}, and the central region, which can be associated with the forming BCG, the {\it protocluster core}. Note that we will \textit{not} attempt to find protocluster candidates by associations of galaxies, as observer would do. We would suggest to call these {\it protocluster associations}.

In this study, we will use one of the largest fully baryonic cosmological hydrodynamical simulation volumes from the {\it Magneticum} pathfinder simulation suite, which we will introduce in Sec.~\ref{sec:sims}, to identify for the first time protocluster core counterparts to those observed at $z\approx4$ and study their properties in Sec.~\ref{sec:properties}, including member star formation rates and quiescent fractions. Using the full power of the simulations, we will track those protocluster cores and their galaxies down to $z=0$ in Sec.~\ref{sec:track}, also studying the impact of the cosmological parameters as well as the importance of simulation volumes in finding overdensities as massive as those currently observed, predicting maximum observable virialized masses up to $z=10$. Finally, in Sec.~\ref{sec:conclusion}, we will summarize and discuss the results.

\section{The Magneticum Pathfinder Simulations}\label{sec:sims}
To find protoclusters at high redshift comparable to those observed recently in mass, a large fully baryonic simulation volume is required.
For the major part of this study we use one of the largest volumes from the hydrodynamical cosmological simulation suite {\it Magneticum Pathfinder}\footnote{www.magneticum.org} (Dolag et al., 2021, in prep),
for which the resolution is high enough to resolve galaxies down to baryonic masses of $M_\mathrm{bar} > 10^{10}M_\odot$. 
This simulation, {\it Box~2b} \citep[see][Kimmig et al., in prep.]{bocquet:2016,ragagnin:2019}, has a box-size of $(640~\mathrm{Mpc}/h)^{3}$ and a particle mass resolution of $m_\mathrm{DM} = 6.9\times10^{8} M_{\odot}/h$ and $m_\mathrm{Gas} = 1.4\times10^{8} M_{\odot}/h$ for dark matter and gas, respectively. 
Since each gas particle can spawn up to four stellar particles during its lifetime, the mass of a stellar particle is approximately $m_\mathrm{*} \simeq 3.5\times10^{7} M_{\odot}/h$. 
For dark matter and gas particles the same softening is used, with $\epsilon_\mathrm{DM} = \epsilon_\mathrm{Gas} = 3.75~\mathrm{kpc}/h$, while for the stars a softening of $\epsilon_\mathrm{*}= 2~\mathrm{kpc}/h$ was adopted.
For more details on this specific simulation and its clusters and galaxies at low redshifts, see \citet{remus:2017,lotz:2019,harris:2020,lotz:2021}. 

The Magneticum simulations adapt a WMAP~7 $\Lambda$CDM cosmology \citep{komatsu:2011}, i.e. $\sigma_8=0.809$, $h=0.704$, $\Omega_\mathrm{\Lambda} = 0.728$, $\Omega_\mathrm{m} = 0.272$, and $\Omega_\mathrm{b} = 0.0451$. 
For the initial slope of the power spectrum, a value of $n_\mathrm{s} = 0.963$ is used.
The simulation was performed with an updated version of GADGET-3, including, in addition to various modifications in the formulation of SPH \citep{dolag:2004,dolag:2005,donnert:2013,beck:2015}, modernised
versions of the sub-grid physics, especially with respect to the star formation and metal enrichment descriptions \citep{tornatore:2004,tornatore:2007,wiersma:2009}, and the black hole feedback \citep{fabjan:2010,hirschmann:2014}. 
For more details on the physics included in the Magneticum Pathfinder simulations, we refer the reader to \citet{hirschmann:2014}, \citet{teklu:2015}, and \citet{dolag:2017}.

The data of these simulations up to redshift $z\approx2$ are publicly available on the {\it Cosmological Web Portal}\footnote{https://c2papcosmosim.uc.lrz.de/}, see \citet{ragagnin:2017}.
In general, the main suite of the Magneticum simulations encompasses five different simulation volumes: 
{\it Box~0} with a box-length of $(2688~\mathrm{Mpc}/h)^{3}$, 
{\it Box~1} with a box-length of $(896~\mathrm{Mpc}/h)^{3}$, 
{\it Box~2b} with a box-length of $(640~\mathrm{Mpc}/h)^{3}$, 
{\it Box~3} with a box-length of $(128~\mathrm{Mpc}/h)^{3}$, and
{\it Box~4} with a box-length of $(48~\mathrm{Mpc}/h)^{3}$, all adopting the same physics and cosmology as described above.
In addition, the Magneticum simulations include a set of 15 simulations of the {\it Box~1} volume with a resolution of $2\times1526^3$ particles, adopting different
cosmologies with varying $\sigma_8$, $H_0$, $\Omega_\mathrm{0}$, and $\Omega_\mathrm{b}$. These simulations have been introduced by \citet{singh:2019}.

For all simulations, structures are identified using a modified version of SUBFIND \citep{springel:2001,dolag:2009}. 
Total masses (baryonic and dark matter) are calculated in case of $M_\mathrm{vir}$ from the Top-Hat model, with the virial radius $R_\mathrm{vir}$ the radius of the sphere that would include the virial mass in case the structure would be relaxed and virialized. 
This is, of course, a very poor assumption especially at high redshifts, and thus the virial radius is not necessarily a good prescription of the actual radial distribution of the structure. 

\section{Protoclusters at \lowercase{$\mathrm{z\approx4}$}}\label{sec:properties}
As the recently detected massive protoclusters or protocluster cores are at redshifts of about $z\approx 4$ \citep{miller:2018,oteo:2018,rotermund:2021}, we select our protocluster candidates at a similar redshift snapshot of $z=4.2$.
\begin{figure*}
  \begin{center}
    \includegraphics[width=.9\textwidth]{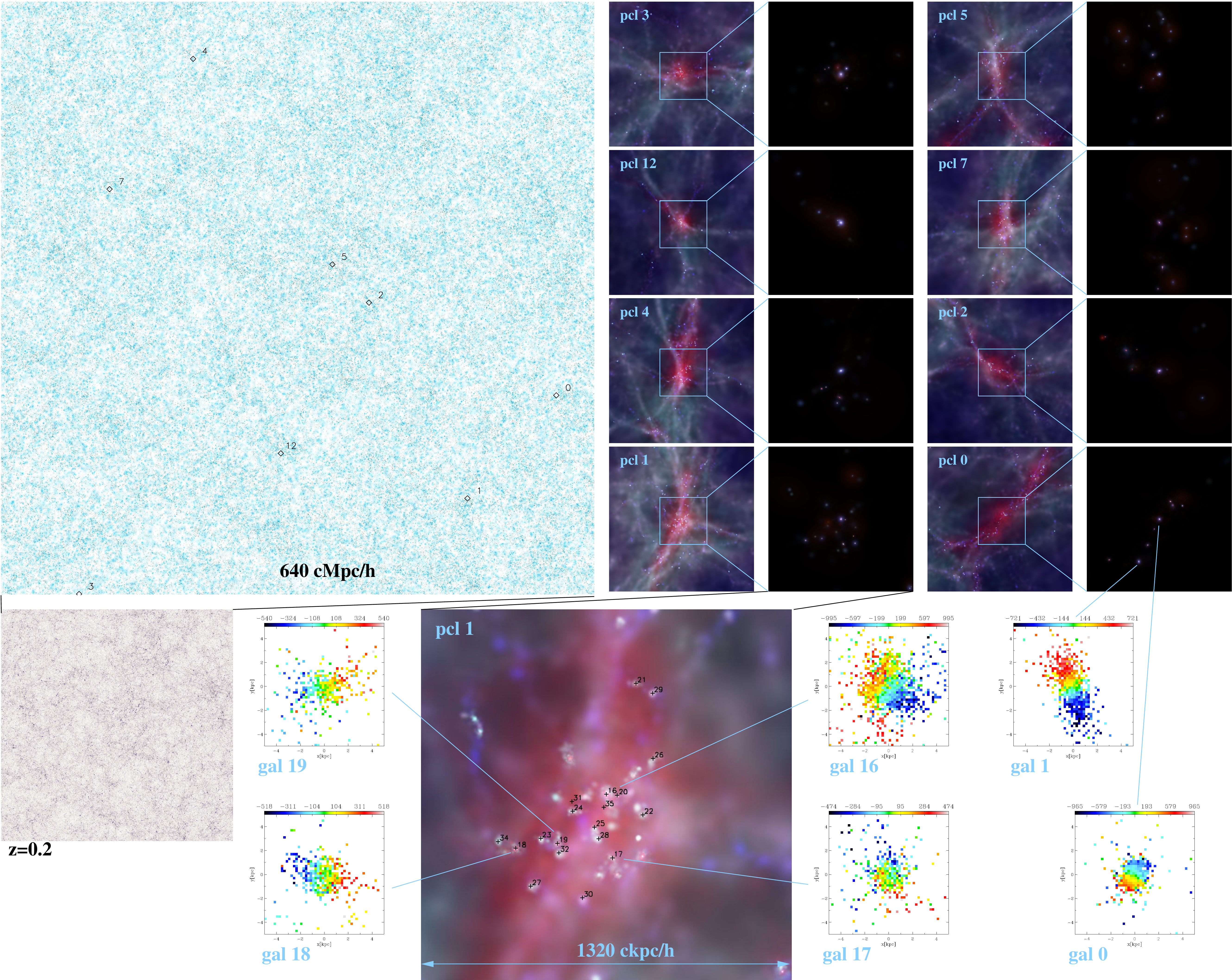}
    \caption{\textit{Upper left:} Stellar distribution of {\it Magneticum Box~2b} at $z=4.2$, with the labeled locations of the protoclusters PCl~3/12/0/5/1/7/2/4.
    \textit{Upper right:} Zoom-in on the 8 selected protoclusters, with the left panels showing the gas content, from cold (blue) to hot (red), and the right panels showing a zoom on the central stellar components within $1320~\mathrm{kpc}/h$ co-moving,
    which corresponds to a physical box length of $353.77~\mathrm{kpc}$, with the colors marking the age of the stars (from young (blue) to old (red)).
    \textit{Lower right:} Larger version of the central area of PCl~1 with both gas and stars shown in the same color scheme as in the small panels, with a box-length of $353.77~\mathrm{kpc}$. 
    The galaxies are labeled according to their stellar mass. 
    In addition, the rotation maps of the gas are shown for six example galaxies, four from PCl~1 and two from PCl~0.
    \textit{Lower left:} Stellar distribution of the {\it Magneticum Box~2b} volume at $z=0$.
    }
  {\label{fig:proto_map}}
\end{center}
\end{figure*}
These protoclusters or protocluster cores consist of several galaxies clustered in a very small volume.
However, it is not known for sure from observations that such structures are already bound, albeit it is very clear that they are already linked together. Therefore, we select our protocluster candidates based on the total mass that is already linked together by a Friends-of-Friends algorithm with a linking length of 0.2. 
At $z=4.2$, this provides us with 42 structures with total masses above $M_\mathrm{tot} = 1\times10^{13}M_\odot$.

However, this is not a possible detection criterion for observations, so we refine our selection critera from this pool of protocluster candidates based on the following four methods that closely mimic observational methods:
First, as the galaxies observed in these protoclusters are in close vicinity and from their velocities they indicate that they are already bound, we assume that these protocluster cores are already bound structures, and thus assume that they are already virialized\footnote{Here we use $M_\mathrm{vir}$, the density contrast calculated from the Top-Hat model. These values (and the ranking) will slightly change if we use $M_\mathrm{200crit}$ or  $M_\mathrm{200mean}$.}.
This delivers the most massive and concentrated structure at that redshift, and we find the most massive system having a total mass of $M_\mathrm{vir} = 2.148\times10^{13}M_\odot$. 
Tab.~\ref{tab:sims2} lists the 16 most massive bound structures in our simulation at $z=4.22$ according to their total (virial) mass (upper part).

A second approach is to identify protoclusters based on an already very massive central galaxy (a proto-BCG), with several smaller galaxies in close vicinity. The lower part of Tab.~\ref{tab:sims2} lists the 16 protocluster candidated with the most massive stellar galaxy $M_\mathrm{CD}$. 
As can be seen immediately from this table, there is no direct correlation between the total mass of the protocluster candidates and the stellar mass of the most massive galaxy in the structure; 
in fact, only half of the structures with the highest total mass are present in the list of 16 structures with the most massive stellar components. 
This already indicates that there is no simple indicator that uniquely links the total mass growth and the growth of the stellar components at this epoch.
\begin{table*}
\caption{Details of the 16 highest-ranked protocluster candidates at $z=4.2$, selected according to their total (dark plus baryonic) mass $M_\mathrm{vir}$ calculated from the Top-Hat model (upper table block), 
and according to the stellar mass of their most massive galaxy $M_\mathrm{CD}$ (lower table block). 
The mass rank refers to the total (dark plus baryonic) mass at the given redshift. $N_\mathrm{gal}$ is the total number of galaxies inside the protocluster, with $N_\mathrm{sf}$ and $N_\mathrm{qui}$ referring to the number of star forming and quiescent cluster member galaxies, respectively, with galaxies with $sSFR < 0.3\times t_\mathrm{Hub}$ defined as quiescent. 
$M_*$ is the total stellar mass inside the virial radius, including all member galaxies as well as the ICM. 
$SFR_\mathrm{total}$ refers to the star formation rate inside the whole protocluster, summed over all galaxy members. 
The last two columns are the virial mass $M_\mathrm{vir}$ and the mass rank of the protoclusters at $z=0$.
The eight protoclusters that we study in more detail in this work are highlighted in gray if they are part of the respective 16 highest-ranks.
}
\label{tab:sims2}
\def\swd{\hphantom{0}}
\def\dwd{\hphantom{00}}
\def\colorrow{\noalign{\color[gray]{0.9}\hrule height 13pt}\\[-26pt]}
\def\hlinecolorrow{\noalign{\color[gray]{0.9}\hrule height 13.3pt}\\[-26.5pt]\hline}
\begin{center}
\begin{tabular}{c | cccccccc | cc}
\hline\hline
ID      & mass rank    & $M_\mathrm{vir}$  & $M_\mathrm{CD}$    & $N_\mathrm{gal}$ & $M_*$               & $SFR_\mathrm{total}$      & $N_\mathrm{sf}$ & $N_\mathrm{qui}$    & $M_\mathrm{vir}$($z=0$) & mass rank \\
        & ($z=4.2$) & $[10^{13}M_\odot]$& $[10^{11}M_\odot]$ &                  & $[10^{11}M_\odot]$  & $[M_\odot/\mathrm{yr}]$   &                 & 			& $[10^{14}M_\odot]$             & ($z=0$) \\
[1pt]\hlinecolorrow
   \dwd 0 & \dwd 0 &  2.148     &  3.668       &          14	  &  11.965         &      2295.31		&          14	  &           0	&    14.332 	&   \swd\swd\swd\swd6	\\
\colorrow
   \dwd 3 & \dwd 1 &  2.006     &  4.995       &          11	  & \swd9.541       &      2858.06		&          10	  &           1	& \swd7.608 	&      \swd\swd\swd67	\\
\colorrow
   \dwd 2 & \dwd 2 &  1.971     &  5.661       &     \swd  9	  & \swd9.536       &      1511.79		&       \swd8	  &           1	& \swd3.049 	&         \swd\swd776	\\
\colorrow
   \dwd 4 & \dwd 3 &  1.954     &  8.553       &          10	  & \swd9.767       &      1664.15		&          10	  &           0	& \swd4.229 	&         \swd\swd341	\\
\colorrow
   \dwd 1 & \dwd 4 &  1.702     &  5.735       &          19	  & \swd9.278       &      1875.30		&          17	  &           2	&    10.156 	&      \swd\swd\swd32	\\
   \swd13 & \dwd 5 &  1.568     &  4.452       &     \swd  9	  & \swd6.624       &      1937.18		&       \swd9	  &           0	& \swd5.269 	&         \swd\swd207	\\
   \dwd 8 & \dwd 6 &  1.565     &  3.868       &          13	  & \swd8.925       &      1918.91		&          12	  &           1	& \swd3.185 	&         \swd\swd702	\\
\colorrow
   \swd12 & \dwd 7 &  1.543     &  5.823       &     \swd  6	  & \swd7.053       &      2651.89		&       \swd6	  &           0	& \swd1.334 	&            \swd3666	\\
   \dwd 6 & \dwd 8 &  1.473     &  5.713       &     \swd  8	  & \swd9.908       &      1751.19		&       \swd6	  &           2	& \swd6.530 	&         \swd\swd106	\\
   \swd14 & \dwd 9 &  1.445     &  4.640       &     \swd  7	  & \swd7.695       &      1377.81		&       \swd5	  &           2	& \swd5.623 	&         \swd\swd165	\\
   \swd11 & \swd10 &  1.422     &  7.353       &     \swd  7	  & \swd8.717       &      1959.89		&       \swd6	  &           1	& \swd2.291 	&            \swd1364	\\
   \swd18 & \swd11 &  1.394     &  4.608       &          15	  & \swd6.136       &      1525.29		&          13	  &           2	&    10.727 	&      \swd\swd\swd25	\\
   \swd26 & \swd12 &  1.381     &  4.786       &     \swd  9	  & \swd6.339       &      1205.19		&       \swd9	  &           0	& \swd0.324 	&               27831	\\
   \swd21 & \swd13 &  1.359     &  6.354       &     \swd  8	  & \swd7.034       &   \swd948.03		&       \swd4	  &           4	&    13.964 	&      \swd\swd\swd10	\\
   \swd15 & \swd14 &  1.358     &  7.409       &     \swd  7	  & \swd7.754       &      1419.14		&       \swd6	  &           1	& \swd2.072 	&            \swd1665	\\
   \swd23 & \swd15 &  1.349     &  3.849       &     \swd  9	  & \swd5.711       &      1809.63		&       \swd8	  &           1	& \swd3.832 	&         \swd\swd447	\\
[1pt]\hlinecolorrow
   \dwd 4 & \dwd 3 &  1.954     &  8.553       &          10	  & \swd9.767       &      1664.15		&          10	  &           0	& \swd4.229 	&         \swd\swd341	\\
   \swd16 & \swd17 &  1.299     &  8.404       &     \swd  6	  & \swd8.450       &      1335.22		&       \swd3	  &           3	& \swd1.476 	&            \swd3073	\\
   \swd15 & \swd14 &  1.358     &  7.409       &     \swd  7	  & \swd7.754       &      1419.14		&       \swd6	  &           1	& \swd2.072 	&            \swd1665	\\
   \swd11 & \swd10 &  1.422     &  7.353       &     \swd  7	  & \swd8.717       &      1959.89		&       \swd6	  &           1	& \swd2.291 	&            \swd1364	\\
   \swd25 & \swd18 &  1.250     &  7.017       &     \swd  3	  & \swd7.479       &      1452.07		&       \swd3	  &           0	&    11.603 	&      \swd\swd\swd18	\\
   \swd21 & \swd13 &  1.359     &  6.354       &     \swd  8	  & \swd7.034       &   \swd948.03		&       \swd4	  &           4	&    13.964 	&      \swd\swd\swd10	\\
   \swd63 & \swd46 &  0.950     &  6.184       &     \swd  2	  & \swd6.215       &      1023.38		&       \swd1	  &           1	& \swd1.156 	&            \swd4634	\\
   \swd36 & \swd31 &  1.090     &  5.849       &     \swd  7	  & \swd5.983       &      1383.49		&       \swd6	  &           1	& \swd3.973 	&         \swd\swd409	\\
   \swd12 & \dwd 7 &  1.543     &  5.823       &     \swd  6	  & \swd7.053       &      2651.89		&       \swd6	  &           0	& \swd1.334 	&            \swd3666	\\
   \swd70 & \swd52 &  0.922     &  5.763       &     \swd  2	  & \swd5.769       &      1148.47		&       \swd1	  &           1	& \swd1.504 	&            \swd2967	\\
\colorrow
   \dwd 1 & \dwd 4 &  1.702     &  5.735       &          19	  & \swd9.278       &      1875.30		&          17	  &           2	&    10.156 	&      \swd\swd\swd32	\\
   \dwd 6 & \dwd 8 &  1.473     &  5.713       &     \swd  8	  & \swd9.908       &      1751.19		&       \swd6	  &           2	& \swd6.530 	&         \swd\swd106	\\
\colorrow
   \dwd 2 & \dwd 2 &  1.971     &  5.661       &     \swd  9	  & \swd9.536       &      1511.79		&       \swd8	  &           1	& \swd3.049 	&         \swd\swd776	\\
   \swd69 & \swd54 &  0.916     &  5.655       &     \swd  2	  & \swd5.762       &   \swd919.90		&       \swd2	  &           0	& \swd4.412 	&         \swd\swd313	\\
   \swd49 & \swd34 &  1.064     &  5.603       &     \swd  4	  & \swd5.884       &      1408.18		&       \swd4	  &           0	& \swd2.984 	&         \swd\swd802	\\
   \swd34 & \swd25 &  1.163     &  5.598       &     \swd  6	  & \swd6.200       &      1181.58		&       \swd5	  &           1	& \swd1.150 	&            \swd4668	\\
\hline                                                                                                                                                                                
\end{tabular}
\end{center}
\end{table*}

\begin{table*}
\caption{Same as Tab.~\ref{tab:sims2} but for the 16 highest-ranking protocluster candidates at $z=4.2$ selected according to their total star formation rate $\mathrm{sfr}_\mathrm{total}$ (upper table block), and their total number of member galaxies $N_\mathrm{gal}$ (lower table block).}
\label{tab:sims1}
\def\swd{\hphantom{0}}
\def\dwd{\hphantom{00}}
\def\colorrow{\noalign{\color[gray]{0.9}\hrule height 13pt}\\[-26pt]}
\def\hlinecolorrow{\noalign{\color[gray]{0.9}\hrule height 13.3pt}\\[-26.5pt]\hline}
\begin{center}
\begin{tabular}{c | cccccccc | cc}
\hline\hline
ID 	& mass rank    & $M_\mathrm{vir}$  & $M_\mathrm{CD}$    & $N_\mathrm{gal}$ & $M_*$ 	            & SFR$_\mathrm{total}$ 	& $N_\mathrm{sf}$ & $N_\mathrm{qui}$ 	& $M_\mathrm{vir}$($z=0$) & mass rank \\
   	& ($z=4.2$) & $[10^{13}M_\odot]$& $[10^{11}M_\odot]$ &		      & $[10^{11}M_\odot]$  & $[M_\odot/\mathrm{yr}]$   & 	          & 			& $10^{14}[M_\odot]$  		  & ($z=0$) \\
[1pt]\hlinecolorrow
   \dwd3  & \dwd 1  &  2.006     &  4.995       &          11     & \swd9.541	    &      2858.06              &          10     &           1	& \swd7.608 	&      \swd\swd\swd67	\\
\colorrow
   \swd12 & \dwd 7  &  1.543     &  5.823       &      \swd 6     & \swd7.053	    &      2651.89              &      \swd 6     &           0	& \swd1.334 	&            \swd3666	\\
\colorrow
   \dwd0  & \dwd 0  &  2.148     &  3.668       &          14     &    11.965	    &      2295.31              &          14     &           0	&    14.332 	&   \swd\swd\swd\swd6	\\
\colorrow
   \dwd5  & \swd43  &  0.983     &  2.261       &          25     & \swd7.341	    &      2115.63              &          25     &           0	&    10.742 	&      \swd\swd\swd24	\\
     141  & \swd88  &  0.799     &  3.902       &      \swd 6     & \swd4.170	    &      2061.60              &      \swd 5     &           1	& \swd4.421 	&         \swd\swd311	\\
     157  &    124  &  0.743     &  3.787       &      \swd 6     & \swd3.947	    &      1999.84              &      \swd 5     &           1	& \swd0.765 	&            \swd8708	\\
   \swd10 & \swd30  &  1.113     &  2.746       &          14     & \swd7.568	    &      1985.91              &          13     &           1	& \swd8.756 	&      \swd\swd\swd44	\\
   \swd11 & \swd10  &  1.422     &  7.353       &      \swd 7     & \swd8.717	    &      1959.89              &      \swd 6     &           1	& \swd2.291 	&            \swd1364	\\
   \swd13 & \dwd 5  &  1.568     &  4.452       &      \swd 9     & \swd6.624	    &      1937.18              &      \swd 9     &           0	& \swd5.269 	&         \swd\swd207	\\
   \dwd 8 & \dwd 6  &  1.565     &  3.868       &          13     & \swd8.925	    &      1918.91              &          12     &           1	& \swd3.185 	&         \swd\swd702	\\
   \swd64 & \swd37  &  1.030     &  4.177       &      \swd 4     & \swd4.581	    &      1916.37              &      \swd 3     &           1	& \swd1.464 	&            \swd3124	\\
\colorrow
   \dwd 1 & \dwd 4  &  1.702     &  5.735       &          19     & \swd9.278	    &      1875.30              &          17     &           2	&    10.156 	&      \swd\swd\swd32	\\
   \swd85 &    183  &  0.677     &  2.130       &      \swd 7     & \swd4.282	    &      1811.17              &      \swd 6     &           1	& \swd0.819 	&            \swd7864	\\
   \swd23 & \swd15  &  1.349     &  3.849       &      \swd 9     & \swd5.711	    &      1809.63              &      \swd 8     &           1	& \swd3.832 	&         \swd\swd447	\\
   \swd24 & \swd21  &  1.225     &  3.337       &          10     & \swd5.476	    &      1789.22              &      \swd 9     &           1	&    11.655 	&      \swd\swd\swd17	\\
   \dwd 6 & \dwd 8  &  1.473     &  5.713       &      \swd 8     & \swd9.908	    &      1751.19              &      \swd 6     &           2	& \swd6.530 	&         \swd\swd106	\\
[1pt]\hlinecolorrow
   \dwd 5 & \swd 43 &  0.983     &  2.261       &          25	  & \swd7.341       &      2115.63		&          25	  &           0	&    10.742 	&      \swd\swd\swd24	\\
\colorrow
   \dwd 1 & \dwd  4 &  1.702     &  5.735       &          19	  & \swd9.278       &      1875.30		&          17	  &           2	&    10.156 	&      \swd\swd\swd32	\\
\colorrow
   \dwd 7 & \swd 92 &  0.792     &  4.498       &          18	  & \swd8.507       &      1484.96		&          16	  &           2	&    10.344 	&      \swd\swd\swd29	\\
   \swd54 &     304 &  0.598     &  2.378       &          16	  & \swd4.512       &      1349.63		&          16	  &           0	& \swd6.042 	&         \swd\swd130	\\
   \swd42 & \swd 82 &  0.821     &  1.588       &          16	  & \swd4.301       &      1264.03		&          14	  &           2	&    15.947 	&   \swd\swd\swd\swd3	\\
   \swd45 & \swd 59 &  0.897     &  1.561       &          16	  & \swd4.760       &      1095.77		&          15	  &           1	&    16.091 	&   \swd\swd\swd\swd2	\\
   \swd32 & \swd 51 &  0.923     &  0.888       &          16	  & \swd5.162       &      1294.38		&          15	  &           1	& \swd3.388 	&         \swd\swd610	\\
   \swd19 & \swd 36 &  1.030     &  4.999       &          15	  & \swd6.304       &      1525.21		&          15	  &           0	& \swd3.631 	&         \swd\swd519	\\
   \swd20 & \swd 29 &  1.120     &  2.322       &          15	  & \swd5.413       &      1390.69		&          13	  &           2	& \swd6.279 	&         \swd\swd115	\\
   \swd18 & \swd 11 &  1.394     &  4.607       &          15	  & \swd6.136       &      1525.29		&          13	  &           2	&    10.727 	&      \swd\swd\swd25	\\
   \swd56 &     286 &  0.607     &  1.534       &          14	  & \swd4.393       &   \swd990.11		&          14	  &           0	& \swd0.791 	&            \swd8280	\\
   \swd10 & \swd 30 &  1.113     &  2.746       &          14	  & \swd7.568       &      1985.91		&          13	  &           1	& \swd8.756 	&      \swd\swd\swd44	\\
\colorrow
   \dwd 0 & \dwd  0 &  2.148     &  3.668       &          14	  &  11.965         &      2295.31		&          14	  &           0	&    14.332 	&   \swd\swd\swd\swd6	\\
      110 &     466 &  0.529     &  0.561       &          13	  & \swd3.892       &      1023.11		&          13	  &           0	& \swd2.940 	&         \swd\swd821	\\
   \swd30 &     172 &  0.686     &  2.465       &          13	  & \swd5.481       &      1192.98		&          13	  &           0	& \swd3.657 	&         \swd\swd508	\\
   \swd66 &     110 &  0.760     &  1.806       &          13	  & \swd4.215       &      1175.05		&          12	  &           1	& \swd1.659 	&            \swd2516	\\
\hline
\end{tabular}
\end{center}
\end{table*}
However, observationally the total (or dynamical) mass is difficult to measure, and structures at high redshift are usually detected due to their extreme luminosity indicative of high amounts of star formation, or due to their overdensity in number of galaxies. 
Therefore, with method 3 we also select 16 protocluster candidates as the most star forming structures at $z=4.2$ (see upper part of Tab.~\ref{tab:sims1}). 
Interestingly, the structure with the highest star formation rate of $\mathrm{sfr} = 2858.06~M_\odot/yr$ is the second most massive structure in our simulation box, and the three most star forming structures are all among the 16 most massive structures. 
However, similar to what we found for the stellar mass of the most massive galaxy, about half of the most star forming structures are not among the most massive structures in our sample.

Additionally, we also select the 16 structures with the highest number of galaxies at $z=4.2$ (see lower part of Tab.~\ref{tab:sims1}), with the richest structure hosting 25 galaxies, all of which are star forming. 
This structure has, in fact, a total mass of only $M_\mathrm{vir} = 9.83\times10^{12}M_\odot$, and thus is definitely not among the 42 most massive structures in our simulation at $z=4.2$. 
Here, we find a first very interesting tendency: From the 16 richest structures, only three are among the 16 most massive structures in our simulation box, clearly indicating that richness in member numbers is not a good tracer for the most massive structures at redshifts as high as $z=4.2$

\subsection{Selecting a set of protoclusters at $z=4.2$}
From each of the four categories introduced above, we now choose a total of 8 protoclusters to study in more detail in this work, which we will call protoclusters (PCl) in the following and which are highlighted in gray in tables \ref{tab:sims2} and \ref{tab:sims1}: 
PCl~0, PCl~1, PCl~2, PCl~3, PCl~4, PCl~5, PCl~7, and PCl~12.
Images of all these 8 clusters can be seen in the upper right panels of Fig.~\ref{fig:proto_map}, with each left panel showing the gas of the protocluster and its environment, and the right panel showing the stellar particles in the innermost $1329~\mathrm{kpc}/h$ comoving (which is $353.77~\mathrm{kpc}$ physical at $z=4.2$), centered around the protocluster core.
In particular, those 8 clusters were chosen as follows:
PCl~3, PCl~12, PCl~0, and PCl~5 are the four protoclusters with the highest star formation rates, PCl~1 also being among the 16 protoclusters with the highest star formation rates (rank 11);
PCl~5, PCl~1, and PCl~7 are the three richest protoclusters in terms of galaxy membership, with PCl~0 also being among the 16 richest protoclusters (rank 12);
PCl~0, PCl~3, PCl~2, PCl~4, and PCl~1 are the most massive protoclusters with respect to the total mass, with PCl~12 being also among the 16 most massive ones (rank 7);
Finally, PCl~4 is the structure with the most massive stellar galaxy, with the central galaxies of PCl~1 and PCl~2 also belonging to the 16 most massive central galaxies (rank 10 and 12, respectively).

PCl~1 is the only protocluster that is part of all four categories, and its innermost part is shown in the lower central panel of Fig.~\ref{fig:proto_map} with both stars and gas, with gas rotation maps shown for four of its most gas-rich galaxy members.
Interestingly, the properties of our PCl~1 are strikingly similar to those found for {\it SPT2349-56} \citep{miller:2018,rotermund:2021}: 
PCl~1 has a BCG stellar mass of $M_\mathrm{CD} = 5.7\times10^{11}M_\odot$, while the BCG found for {\it SPT2349-56} has been reported to have $M_\mathrm{BCG} = 3.2\times10^{11}M_\odot$; 
PCl~1 has 19 member galaxies above $M_\mathrm{bar} > 10^{10}M_\odot$, while there are so far 14 member galaxies reported for {\it SPT2349-56}.
And the total stellar mass reported for the central part of {\it SPT2349-56} is about $M_\mathrm{*} = 1.2\times10^{12}M_\odot$, while the total stellar mass in the virialized region of PCl~1 is $M_\mathrm{*} = 9.3\times10^{11}M_\odot$.
Gas rotation maps are also shown for two gas-rich galaxies from PCl~0, which is part of three of our four protocluster candidate selection criteria.

The cold gas disks in those gas-rich galaxies found in our protoclusters are of similar gas masses as those observed at high redshifts \citep[e.g.,][]{dannerbauer:2017}, and are typically all rotating at rather high velocities of about 200-600~km/s, which is slightly higher than the values reported from observations so far \citep{smit:2018,jones:2021}, albeit these observed galaxies are not inside protoclusters but rather identified due to their high UV luminosity, and thus they are not directly comparable to the galaxies inside protoclusters shown in this work.
As can be seen from the upper left panel of Fig.~\ref{fig:proto_map}, all of the protoclusters sit at knots of the cosmic filaments, having already a heated atmosphere and accrete galaxies and mostly cold gas along the filaments, which penetrate deeply into the hot atmosphere. 
A detailed analysis of the hot atmosphere of these structures is outside the scope of this paper, where we will mainly focus on the properties of the galaxies and the growth of the structures, however, we will quickly discuss some aspects of this in Sec.~\ref{sec:ICMprop}.
Still, as can be seen in the individual panels of Fig.~\ref{fig:proto_map}, the geometry of these clusters can be quite different, for example PCl~0 shows a quite striking linear geometry of the main member galaxies, while PCl~12 has a more spherical geometry.

\subsection{Properties of the galaxies within protoclusters~at~$z=4.2$}
\begin{figure}
  \begin{center}
    \includegraphics[width = .9\columnwidth]{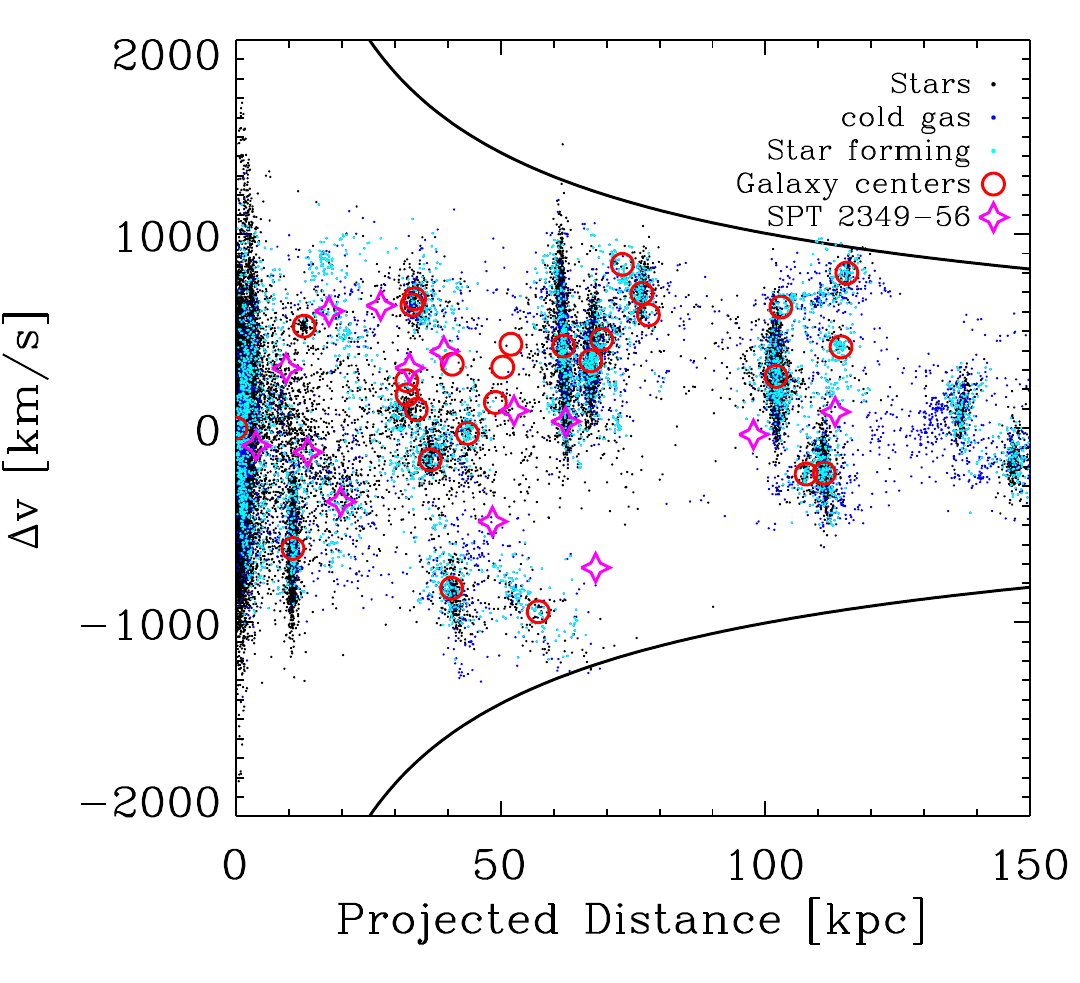}
    \caption{Comparison of the phase-space distribution of the galaxies within {\it SPT2349-56} (pink open stars) and within our protocluster PCL~1 (red open circles for the centres of mass of the member galaxies).
      In addition, black small dots mark the stars, blue small dots mark the cold gas particles,  and cyan small dots are star-forming gas particles in the individual member galaxies.
      Black lines mark the escape velocity assuming a relaxed NFW halo for our protocluster PCL~1.
      }
  {\label{fig:phase_space}}
  \end{center}
\end{figure}
The eight protoclusters selected from the simulation can clearly be classified as already bound systems.
Note that by construction our selected protoclusters always have to be bound systems, and we can assume that the halo finding based on {\it SubFind} identifies all such systems in the simulations. 
Here, we do not intend to find systems which due to projection effects or observational uncertainties could be accidentally identified as spatially close and bound systems although they are not. 
However, as was shown already in Fig.~\ref{fig:proto_map}, some of our selected systems have rather linear geometry of the main member galaxies, like PCl~0 but also PCl~1, clearly indicating active ongoing assembly, similar to what is observed for the surroundings of the massive protocluster structures.
Thus, comparing the phase-space distributions of the member galaxies from our simulated protoclusters to observations still provides important information about the comparability.
Figure \ref{fig:phase_space} shows the phase space distribution of the member galaxies for the protocluster PCl~1 as an example.
As can clearly be seen, the velocities of the individual galaxies, although many of them on first in-fall, are all smaller than the escape velocity. 
Interestingly, the distribution of the individual member galaxies within the phase-space in the simulated protocluster is actually very similar to the distribution reported for {\it SPT2349-56} \citep{miller:2018}, again highlighting that PCl~1 is in fact a good match for {\it SPT2349-56}.

\subsection{Star formation and quiescent fractions in protoclusters at $z=4.2$}
\begin{figure}
  \begin{center}
    \includegraphics[width = \columnwidth]{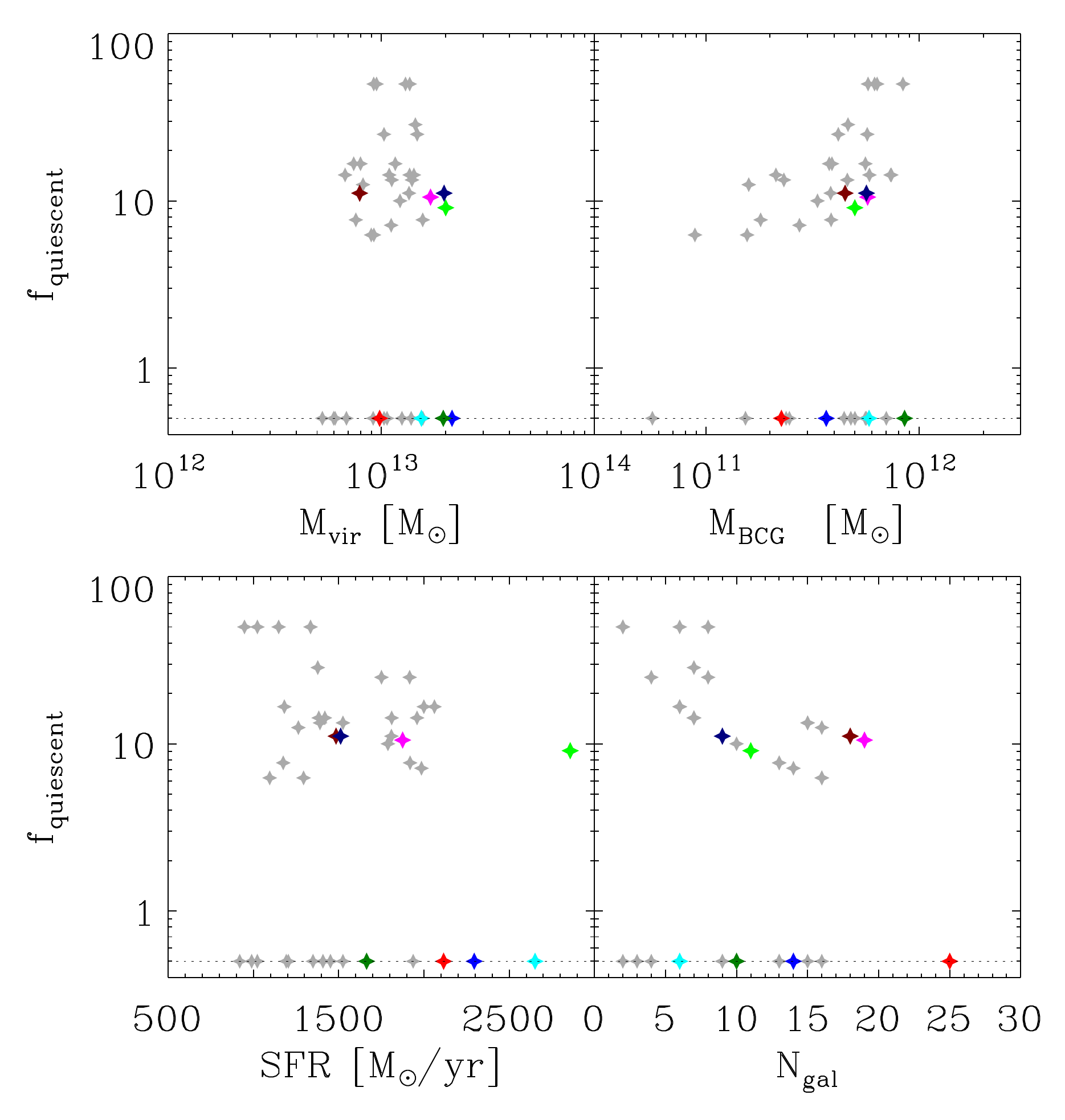}
    \caption{Quiescent fraction for the 42 protocluster candidates versus the four different tracers at $z=4.2$ used to identify protoclusters:
    \textit{Upper left:} Quiescent fraction versus virial mass.
    \textit{Upper right:} Quiescent fraction versus stellar mass of the most massive member galaxy (BCG).
    \textit{Lower left:} Quiescent fraction versus total star formation rate in the virializes region.
    \textit{Lower right:} Quiescent fraction versus richness (number of member galaxies in the virialized region).
    The colored symbols mark the eight specific protoclusters as described in the text, with colors green/\allowbreak cyan/\allowbreak blue/\allowbreak red/\allowbreak magenta/\allowbreak burgundy/\allowbreak darkblue/\allowbreak darkgreen marking PCl~3/12/0/5/1/7/2/4, respectively.
    }
  {\label{fig:redfrac}}
\end{center}
\end{figure}

\begin{figure}
  \begin{center}
    \includegraphics[width = \columnwidth]{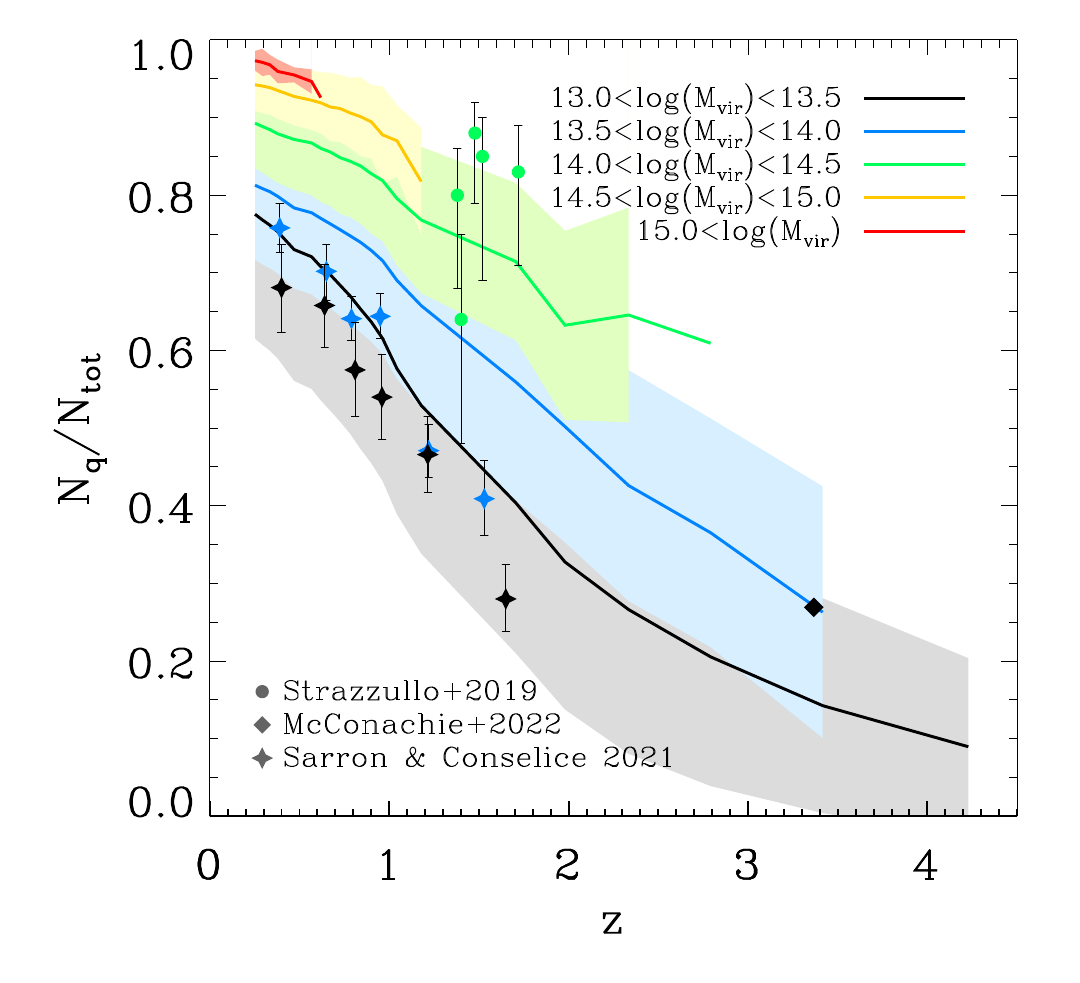}
    \caption{Average fraction of quiescent galaxies with redshift for Magneticum halos of different virial mass ranges, as indicated in the legend.
    Observations of individual clusters from \citet{strazzullo:2019} are included as circles, and the protocluster from \citet{mcconachie:2022} is shown as diamond, with the color indicating the mass range of the observed halos compared to the simulated mass range. 
    In addition, stacked mean quiescent fractions from the {\it Detectivz} survey from \citet{sarron:2021} are included as stars, split into two mass bins corresponding to the two lowest simulated mass ranges shown here (Sarron et al., in prep.).
    }
  {\label{fig:redfracevol}}
\end{center}
\end{figure}

\begin{figure*}
  \begin{center}
    \includegraphics[width=.45\textwidth]{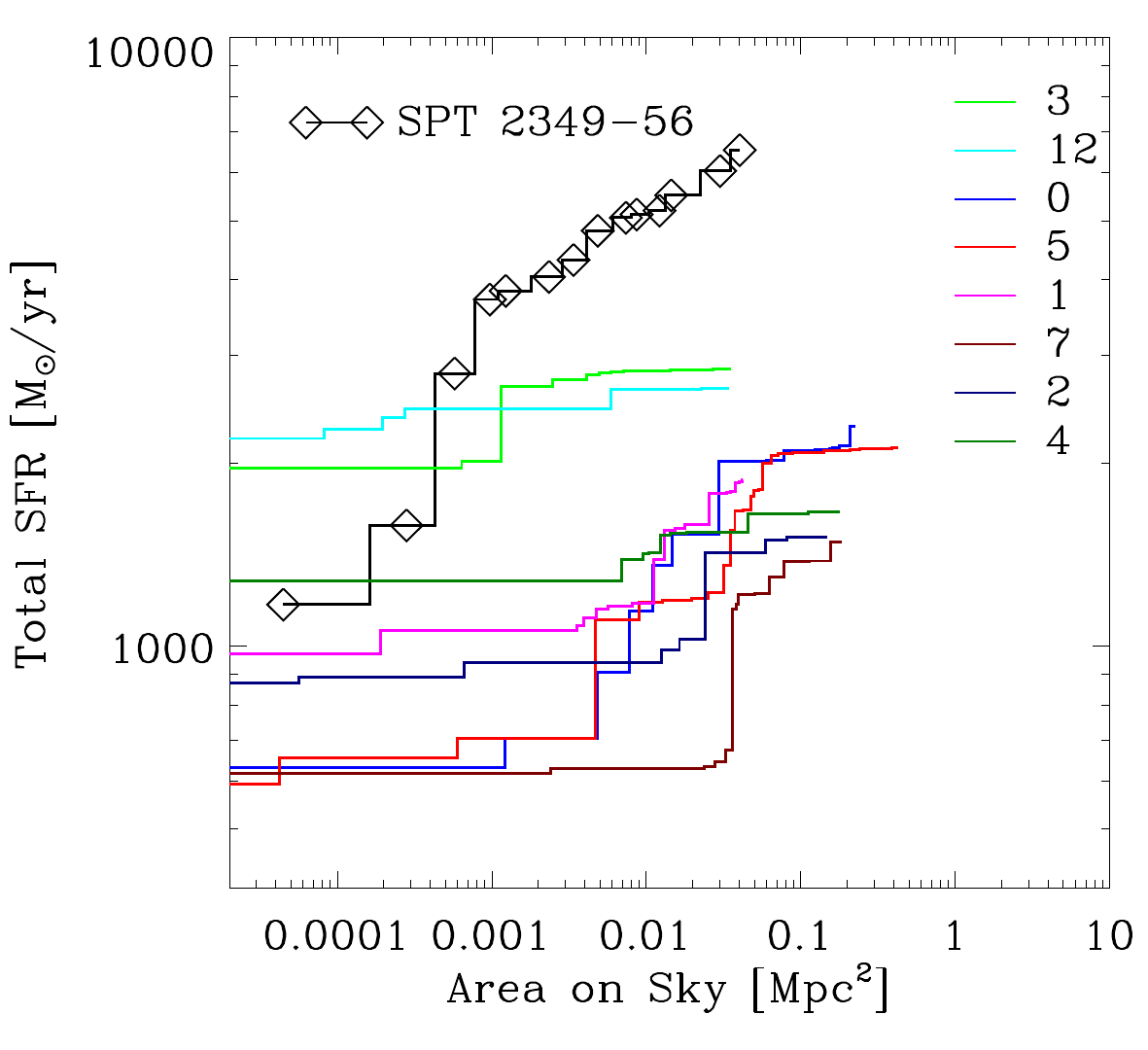}
    \includegraphics[width=.45\textwidth]{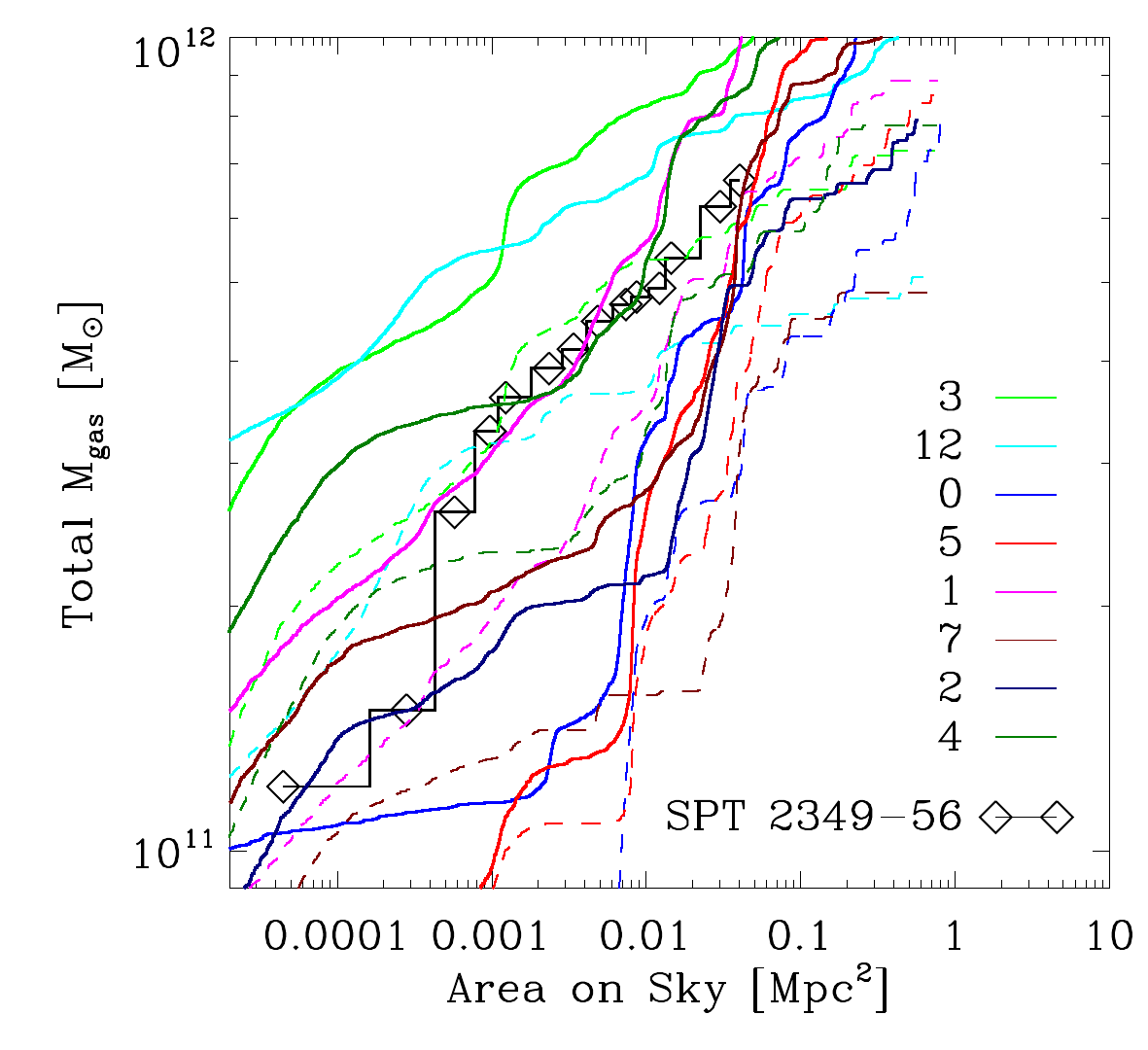}
    \caption{Total star formation rate (left panel) and total gas mass (right panel) versus area-on-sky around the most massive member galaxy for the eight
    example protoclusters PCl~3/12/0/5/1/7/2/4. The solid lines (colors according to the legend) show the data taken directly from the simulation,
    while the dashed lines in the right panel show the total gas mass calculated from the gas inside the galaxies alone.
    For comparison, the values for the protocluster {\it SPT2349-56} at $z=4$ from \citet{miller:2018} are shown as black diamonds and solid line.
    }
  {\label{fig:area_sky_z}}
\end{center}
\end{figure*}

For all our protoclusters we can also distinguish the star forming from the quiescent galaxies. 
Following \citet{franx:2008}, we use the specific star formation rate sSFR, i.e., galaxies with $sSFR < 0.3\times t_\mathrm{Hub}$ are called quiescent, while galaxies with $sSFR > 0.3\times t_\mathrm{Hub}$ are classified as star forming.

\subsubsection{Quiescent Fractions}
As can be seen from Tab.~\ref{tab:sims2} and Tab.~\ref{tab:sims1}, about half of our eight example protoclusters (gray shaded) have no quiescent galaxies yet, while the other half host already one or two quiescent galaxies. 
Interestingly, the appearance of already quenched galaxies is not related to the virial mass of the system or the amount of member galaxies or the star formation rate. 
This gets more clear when looking at all 42 protoclusters, as shown in Fig.~\ref{fig:redfrac}. 
Here, also no trend or correlation with the ranking of the protocluster according to the different selection criteria and the appearance of quiescent galaxies can be seen. 
For example, PCl~5, the protocluster with the highest number of galaxies, has no quiescent galaxy at all, while the protocluster with the largest number of quiescent galaxies, PCl~21, has only 8 galaxies of which half are already quenched.
This clearly shows that the quenched fraction at redshifts as high as $z=4.2$ is not depending on cluster mass, and that environmental quenching at this redshift is not a major quenching mechanism yet, in
agreement with previous results that the morphology-density relation only starts to appear around $z=2$ \citep[e.g.,][]{teklu:2017}.

As already demonstrated in earlier work \citep{lotz:2019}, our simulations generally reproduce the observed fraction of quenched galaxies well at $z=0$ \citep{lotz:2019} and at $z=2.7$ \citep{lustig:2022}. 
Fig.~\ref{fig:redfracevol} shows the evolution of the quenched fraction depending on host halo mass for five different host halo mass ranges, from $z=4.2$ to $z=0$.
As expected, the averaged quenched fraction within galaxy clusters (and groups) not only decreases with increasing redshift, but also decreases with halo mass at a fixed redshift. 
For galaxy clusters above $M_\mathrm{vir}=1\times10^{14}M_\odot$, the amount of quenched galaxies agrees well with observed quenched fractions from SPT by \citet{strazzullo:2019} at redshifts up to $z=1.72$, albeit our average quenched fraction is generally slightly lower than the observed values.

For the lower mass end, we compare our results to observations from the {\it Detectivz} survey by \citet{sarron:2021}. This extensive set of observations covers a redshift range up to $z\approx2$, and the galaxy detection limit in stellar mass is at $M_\mathrm{*}\approx2\times10^{10}M_\odot$, in good agreement with what we can resolve in our simulations, which enables a comparison with respect to the relative quenched fractions. However, virial masses are not measured directly and need to be modelled, which is why the split into halo masses was done based on the stellar masses and not the halo masses, with the groups sorted into the halo mass bin of $5\times10^{13}M_\odot<M_\mathrm{vir}<1\times10^{14}M_\odot$ being selected as having stellar masses of $M_\mathrm{*}>5\times10^{11}M_\odot$, and those that were sorted into the halo mass bin of $1\times10^{13}M_\odot<M_\mathrm{vir}<5\times10^{13}M_\odot$ being selected as having stellar masses of $1\times10^{11}M_\odot<M_\mathrm{*}<5\times10^{11}M_\odot$ (private communication, Sarron et al., in prep.).
As can be seen from the stars in Fig.~\ref{fig:redfracevol}, the predicted quiescent fractions from the simulations are generally larger than what is observed for the group regime, albeit the lower mass bin agrees reasonably within the errorbars. Whether this is due to the split been made based on halo versus stellar mass, or due to other reasons is beyond the scope of this study.

At redshifts higher than $z=2$, quiescent fractions are observationally extremely difficult to obtain. Here, we could only include one data point for the protocluster {\it MAGAZ3NE-J0959} from \citet{mcconachie:2022} at $z\approx3.37$, where the total halo mass is also only a rough estimate. Even though this protocluster has most likely a larger than average fraction of quiescent galaxies, it falls well within the upper 1-$\sigma$ range of our predicted quiescent fractions for halos of a total mass of $1\times10^{13}M_\odot<M_\mathrm{vir}<5\times10^{13}M_\odot$, indicating that our quenching mechanisms produces reasonable quenched fractions at high redshifts, albeit it is only a single object to compare so far. Adding further observations to this will enhance our understanding of the relevant quenching mechanisms at high redshifts in the future.

\subsubsection{Star Forming Galaxies in Protoclusters}
One common finding regarding protoclusters is that models and simulations generally struggle to reproduce the large observed star-formation rates (e.g., \citet{saro:2009} at $z=2$, \citet{granato:2015} up to $z=3$, and \citet{lim:2021} up to z=7). 
In the left panel of Fig.~\ref{fig:area_sky_z} we compare the observed integrated star formation rate as function of area on sky of {\it SPT2349-56} from \citet{miller:2018} with our sample of 8 selected protoclusters.
While the BCGs of several of our protoclusters have a star-formation rate which even exceeds the one of the most star-forming galaxy inside the observed protocluster, the sum of all the observed star-formation rates within the same (virial) area is still a factor of $\approx3$ larger than the one we find in the simulation for all our protoclusters, in agreement with the results found by \citet{bassini:2020} for zoom-in simulations of galaxy clusters. 

On the other hand, comparing the amount of available gas between the observations and the simulations, as shown in the right panel of Fig.~\ref{fig:area_sky_z}, clearly shows that our protocluster candidates have very similar integrated cold gas mass values as the observations, and several of our protoclusters even exceed the observed gas mass values at all distances from the center. 
Even if we calculate the available gas only from what is inside the galaxy members of the protoclusters which (mostly) excludes the hot gas component, as shown as dashed lines in the right panel of Fig.~\ref{fig:area_sky_z}, some of the protoclusters still reproduce the observed values.
Interestingly, the cluster closest in behavior to {\it SPT2349-56} is again PCl~1, with a nearly identical growth of cold gas mass with distance.

In general, this discrepancy between the observed and simulated star formation rates even though the observed and simulated cold gas reservoirs are in agreement and that is also reported for other simulations \citep[e.g.,][]{bassini:2020} could have two reasons: Either the simulations in general do not form star efficiently enough to reproduce the observed values, or the stars are building up at this early times in a much more bursty way compared to the rather continuous rate at which stars are currently formed in simulations.

\subsection{Stellar mass function at high redshift}
\begin{figure*}
  \begin{center}
    \includegraphics[width=.9\textwidth]{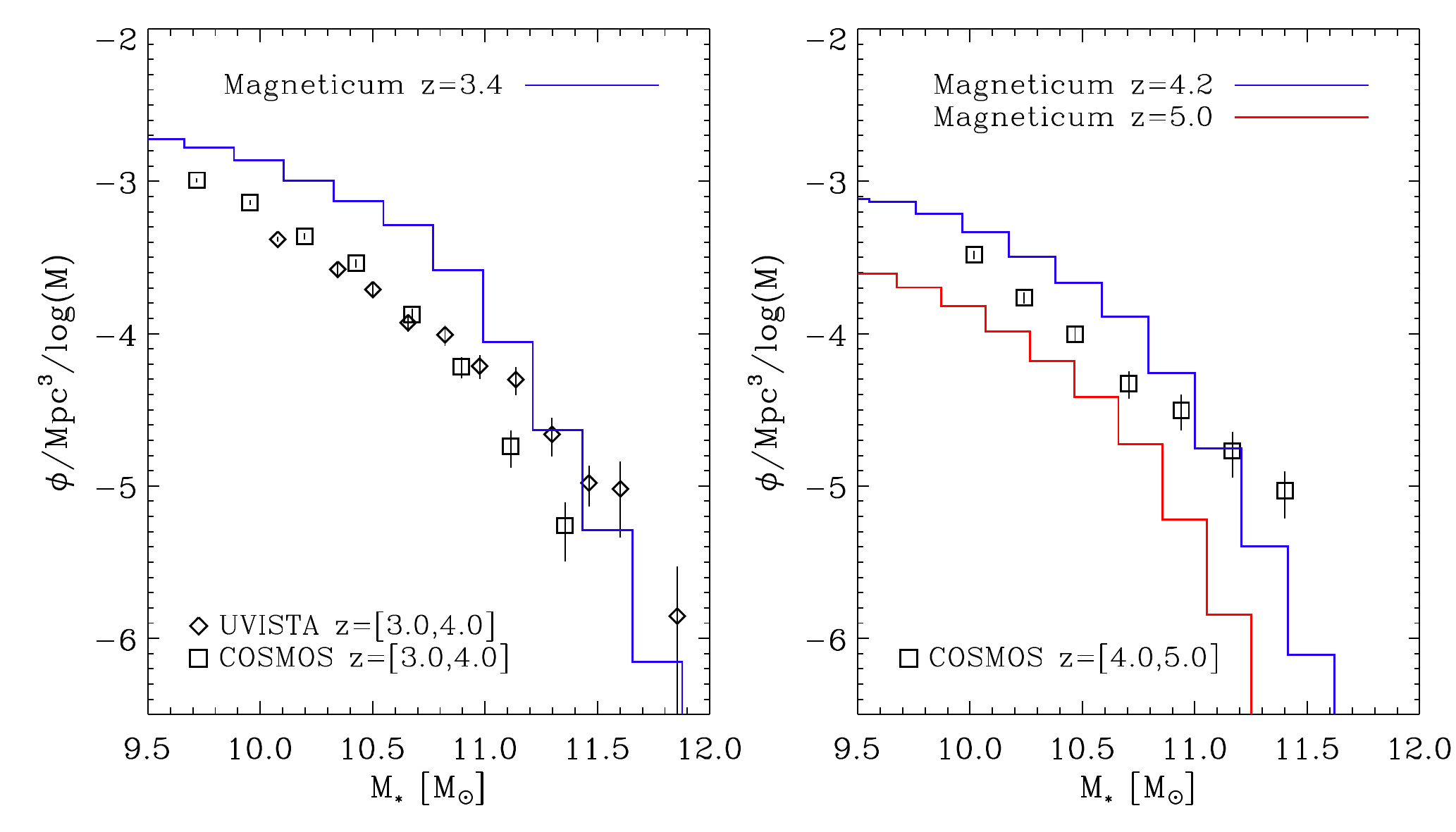}
    \caption{Stellar mass function from the simulation compared to observations from UltraVISTA \citep{ilbert:2013} and COSMOS \citep{davidzon:2017}. \textit{Left panel:} redshift 3--4; \textit{Right panel:} redshift 4-5.}
   {\label{fig:smf_high_z}}
\end{center}
\end{figure*}
One clear test to answer the questions whether the simulations do generally not form stars efficiently enough is to check how the integrated star-formation within the simulation compares to observations, i.e., to compare the stellar mass functions at different redshifts.
For the Magneticum simulations, it has been shown by \citet{hirschmann:2014} and \citet{steinborn:2015} that the stellar mass function is generally well captured between $z=4$ and $z=0$, slightly overshooting the high mass end at $z=0$.
\begin{figure*}
  \begin{center}
    \includegraphics[width=.45\textwidth]{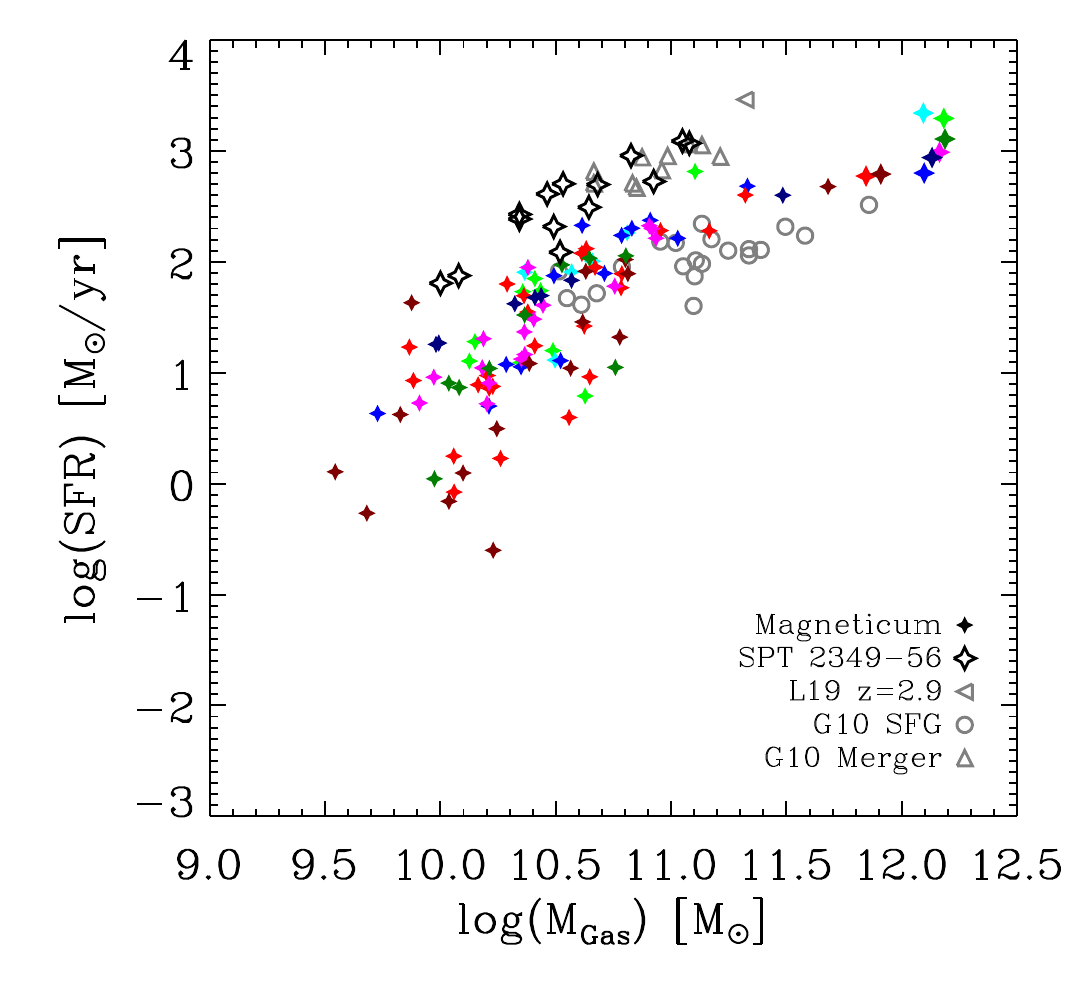}
    \includegraphics[width=.45\textwidth]{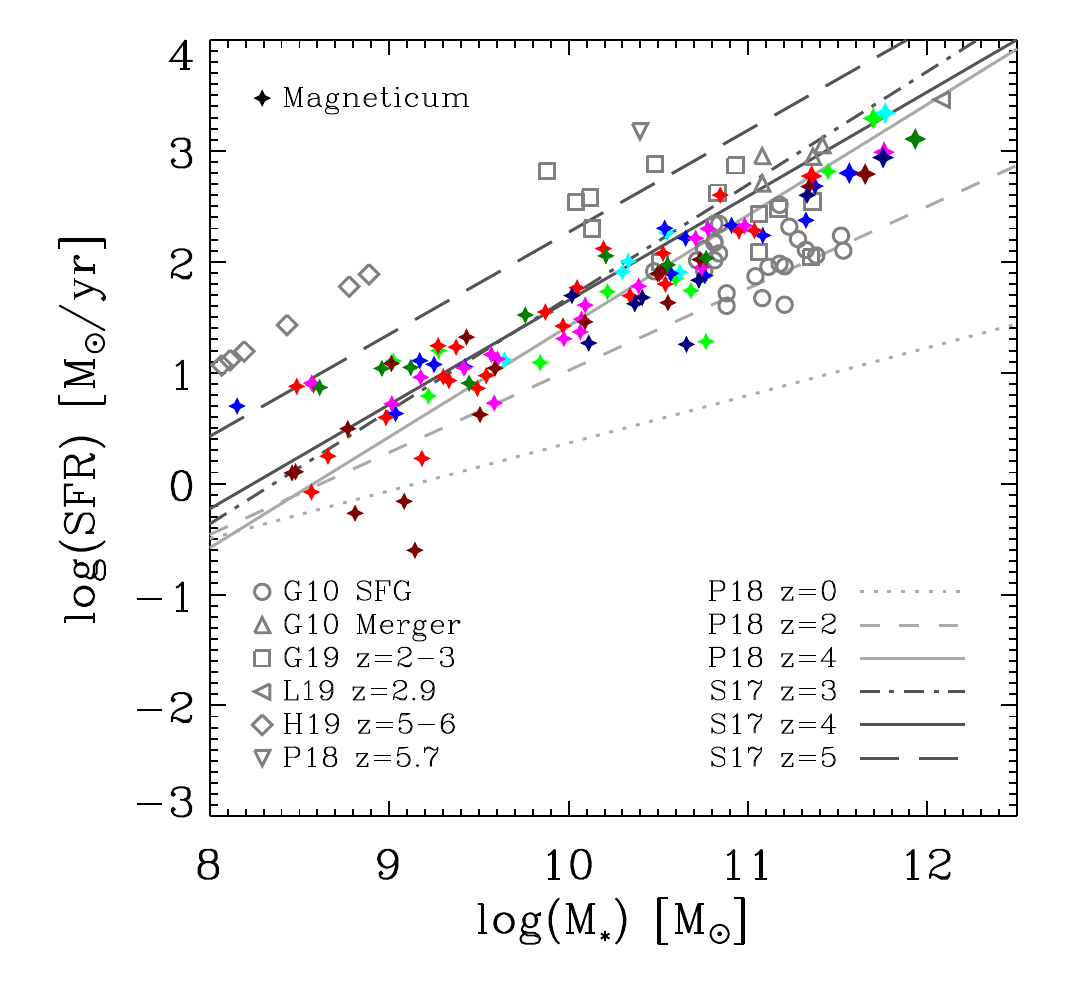}
    \caption{SFR versus $M_\mathrm{Gas}$ (\textit{left panel}) and $M_\mathrm{*}$ (\textit{right panel}).
    Colored stars show the values for the {\it Magneticum} protocluster galaxies for PCl~3/12/0/5/1/7/2/4 at $z=4.2$ (colors as in Fig.~\ref{fig:area_sky_z}).
    For comparison, observational values are included as open symbols:
    In the left panel, the data for the member galaxies from protocluster {\it SPT~2349-56} at $z=4$ from \citet{miller:2018} are shown as open black stars,
    while the gray left-pointing triangle marks the data for the extremely star forming galaxy HXMM05 at $z=2.9$ from \citet{leung:2019}.
    In addition, the normal star forming (gray circles) and merging (gray upward-pointing triangles) galaxies from \citet{genzel:2010} are shown, which
    are at redshifts between $z=1$ and $z=2$.
    In the right panel, the same data points as in the left panel for all but the {\it SPT~2349-56} galaxies are shown with the same symbols.
    In addition, the value for the starbursting galaxy CRLE at $z=5.6$ from \citet{pavesi:2018} is shown as downward-pointing gray triangle, values from
    protocluster members from HELAISS02 ($z=2.2$), HXMM20 ($z=2.6$) and CL~J1001+0220 ($z=2.5$) from \citet{gomez:2019} are shown as gray squares, and member galaxies
    from the most distant spectroscopically confirmed overdensities z57OD ($z=5.7$) and z66OD ($z=6.6$) are shown as gray diamonds.
    Light gray lines show the power-law fits to the star formation main sequence at different redshifts from Herschel \citep{pearson:2018}, while dark gray lines show
    the corresponding fits from Hubble Frontier Fields measurements \citep{santini:2017}.
   }
  {\label{fig:sfr_z}}
\end{center}
\end{figure*}
Fig.~\ref{fig:smf_high_z} shows the stellar mass functions for {\it Box~2b} from $z=3.4$ to $z=5.0$.
Given that (for data storing reasons) the spacing between available outputs of the simulations is relatively large, comparisons with observations, which typically span certain ranges in redshifts, are more difficult. 
Therefore, in the left panel of the figure we show the observations binned in the interval $z=3\dots 4$ as data points, compared to the simulation at $z=3.4$. 
On one hand there is a good agreement at the high mass end, which is the important part for the protoclusters. 
On the other hand, however, the lack of handling AGN feedback in galaxies below a stellar mass of about $10^{10}M_\odot$ due to our resolution limit for the treatment of black holes clearly imprints in an overshooting of the stellar mass function at the low mass end, as a result of slight overcooling at the low mass end. 
The right panel compares the observations binned in the interval $z=4\dots 5$ with the simulation at $z=4.2$ (blue line) and $z=5.0$ (red line). 
Here, the judgment of the agreement between simulations and observations is more difficult, however, given the involved uncertainties in this comparison, the simulations seem to reasonably re-produce the observed stellar mass functions. 
In summary, we conclude that the description of the averaged star-formation within the simulation seems to reasonably well match the real average star formation within the Universe. 
Especially, the difference between the observed and simulated stellar mass functions are far smaller than the observed differences in the star formation rates.
This again supports our speculation that the difference between the simulations and the observations with respect to the star formation rates at high redshifts is caused by the fact that the simulations do not capture processes which lead to a locally, environment dependent higher star formation efficiency, and therefore the simulations are lacking the extreme star bursting systems, while overall producing the right amount of stars.
\begin{figure}
  \begin{center}
    \includegraphics[width=\columnwidth]{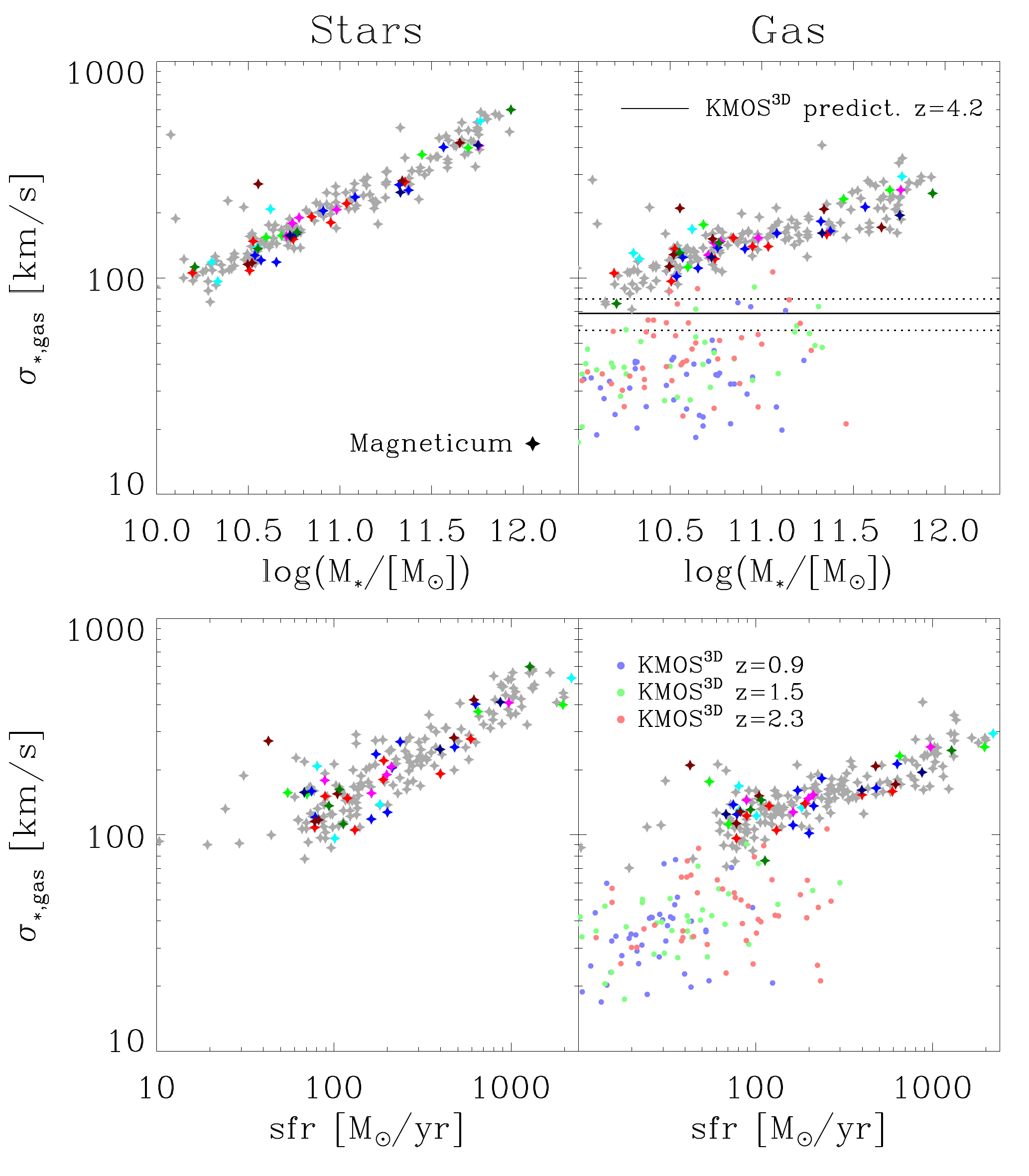}
    \caption{Velocity dispersion within the halfmass radius against stellar mass $M_*$ (\textit{upper panels}) and star formation rate SFR (\textit{lower panels}) for the galaxies in the eight propoclusters colors coded as in Fig.~\ref{fig:area_sky_z},
    and all protocluster candidates from Tab.~\ref{tab:sims2} and Tab.~\ref{tab:sims1} in gray. \textit{Left panels}: for the stellar component of the galaxies. \textit{Right panels}: for the cold gas component of the galaxies.
    Observational data from KMOS$^\textit{3D}$ \citep{uebler:2019} are included in the two right panels as colored dots, redshifts as indicated in the label, together with a prediction of the expected value of the velocity dispersion
    at $z=4.2$, as extrapolated from their Eq.~1.
   }
  {\label{fig:sigma}}
\end{center}
\end{figure}

\subsection{Star formation versus gas reservoir}
The overall star formation of galaxies in the Magneticum simulations follows the observed main sequence of star forming galaxies, both when expressed in terms of stellar mass and in form of gas mass over a large range in redshift, as shown in the right panel of Fig.~\ref{fig:sfr_z}. 
Therefore, the star-formation rate within the galaxies overall are consistent with observations even at the redshift of the protoclusters. However, the simulation misses to reproduce the
population of extremely star-forming galaxies as observed in the protoclusters as well as the star-bursting systems (often classified as mergers) at $z=2$ \citep[e.g.][]{genzel:2010}. 

On the other hand, as can be seen in the left panel of Fig.~\ref{fig:sfr_z}, the star formation rates at a given gas mass are much lower than the observed ones, clearly quantifying the issue already seen from Fig.~\ref{fig:area_sky_z}, that the gas masses agree well with observations while the star fromation rates at the same time are too low.
This again indicates that it is not the general, averaged star-formation description which is insufficient in the simulations but rather the simulations are not producing the short depletion times (or large star formation efficiencies) for systems in special conditions or environments. This results in the general stellar and gas masses fitting well with observations, but the individual star formation rates being too low.

Another quantity intricately connected to the stellar and gas masses as well as the star formation rate is the velocity dispersion in the gas disks where the star formation takes place.
If the gas disks at high redshift are too turbulent compared to observations, with too large velocity dispersions, this could be another reason for the star formation rates to underperform.
As can be seen from the upper panels of Fig.~\ref{fig:sigma}, there are tight correlations in the simulations for the stellar masses and the velocity dispersions found for the stellar (left panel) and the cold gas (right panel) components, repsectively, as well as between the velocity diespersions and the star fromation rate, albeit the correlations between the masses and velocity dispersions is tighter than that found for the velocity dispersions of the components and the star formation rate.
\begin{figure*}
  \begin{center}
    \includegraphics[width = .9\textwidth]{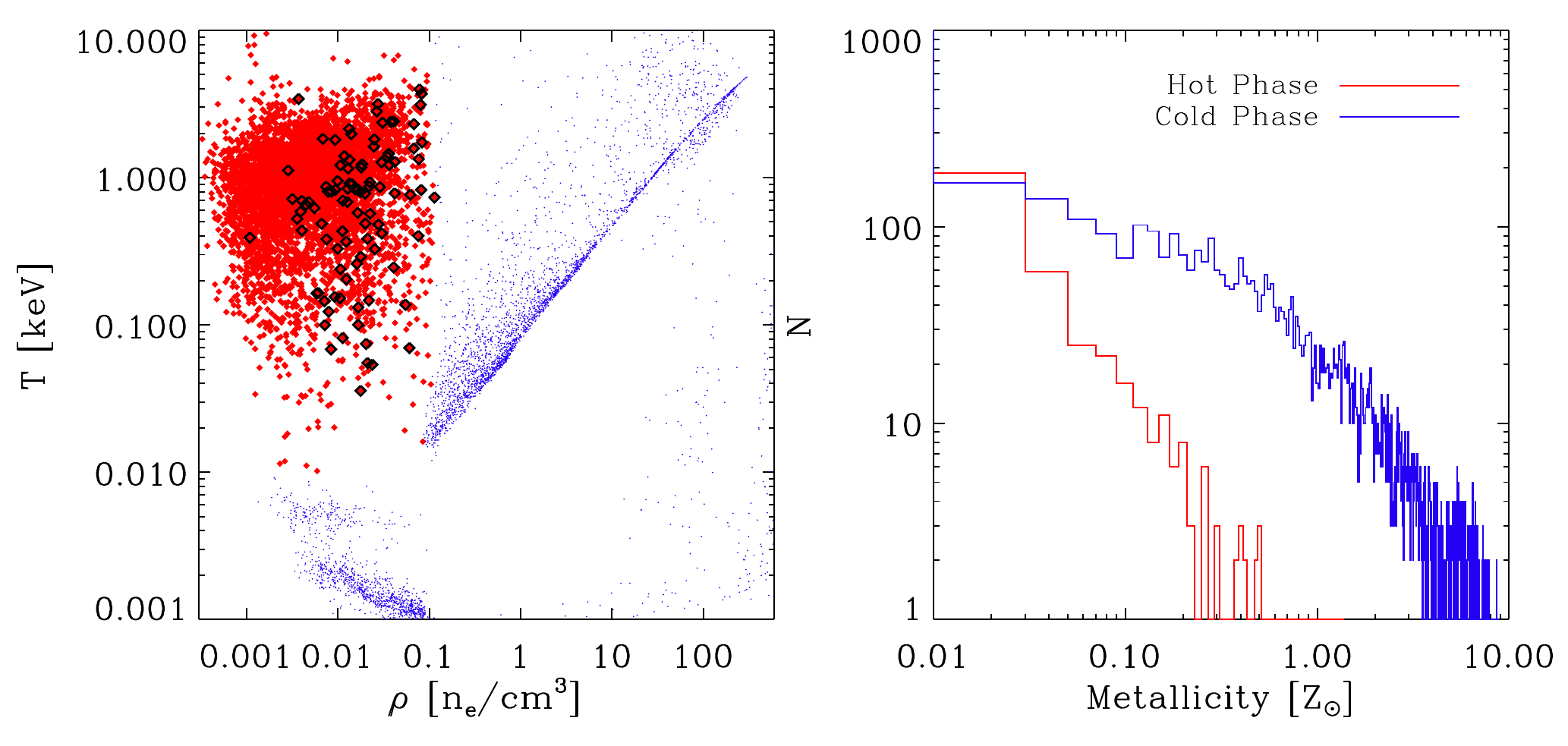}
    \caption{Properties of the gas inside the virial radius of protocluster PCl~1 at $z=4.2$. \textit{Left panel:} Phase diagram where blue dots are the cold, mostly star-forming gas particles, and red points are the hot ICM phase. 
    Black diamonds mark the small subset of the hot ICM which is already chemically enriched above 10\% of the solar value. 
    \textit{Right panel:} Histogram of the chemical enrichment of the cold (blue) and hot (red) phase of the gas.}
  {\label{fig:icm_prop}}
\end{center}
\end{figure*}

While there are no observations of these correlations at redshifts as high as $z=4.2$, velocity dispersions of the cold gas have been observed at redshifts up to $z=2.3$, with predictions for high redshifts from the trends found up to these redshifts \citep{uebler:2019}.
We included these observations in the right panels of Fig.~\ref{fig:sigma} for comparisons of the general trends found for the correlation between the cold gas velocity dispersions and the stellar mass of the galaxies (upper panel) and the star formation rates (lower panel). Both simulations and observations show a correlation between all three quantities with a similar slope, and a general tendency for the gas velocity dispersions to be higher at higher redhsifts. Thus, as far as this comparison is possible, we find general agreement in the trends seen from observations and simulations, however, the absolute velocity dispersions seem to be generally higher than the expected values from the extrapolations of the observations, as indicated by the solid black line in the upper right panel of Fig.~\ref{fig:sigma}.
This enhanced velocity dispersions are generally seen for simulations even down to redshifts of $z=0$, see \citet{vandesande:2019}, and is most likely a result of the differences in particle mass between dark and baryonic masses and the different softenings used for different particle types used in the currently available cosmological simulations \citep{ludlow:2020}, and most prominent for disk galaxies \citep{ludlow:2021}.
How this affects the star formation properties of galaxies in protocluster environments needs to be studies in the future in more detail.
\subsection{Intra Cluster Medium Properties at z=4.2}\label{sec:ICMprop}
So far, we have discussed the properties of the cold gas component of the galaxies found in the protoclusters, however, the build-up of the hot gaseous halo found in galaxy clusters at present day is already taking place in protoclusters at high redshifts. As an example of the expected state of the gas within protoclusters at $z=4.2$, the left panel of Fig.~\ref{fig:icm_prop} shows a phase diagram of the gas within the virial radius of protocluster PCl~1.
At this point, roughly half of the baryonic material within the protocluster is still in the cold phase, which is partially star forming. 
The other half is already in a hot atmosphere, virialized with temperatures centered around $\approx1keV$. 
The presence of gas with temperature of several keV clearly indicates the presence of merger shocks due to the fast growing structures. 
While the cold and star-forming gas is already largely enriched around solar values, the hot intra cluster medium (ICM) is still mainly metal poor, and only $\approx2$\% is already enriched above $10$\% of the solar abundance. Those metal enriched hot gas is generally in denser parts of the hot halo, 
as marked by the black diamonds in the left panel of Fig.~\ref{fig:icm_prop}.
Based on very similar simulations, \citet{biffi:2017,biffi:2018} demonstrated that some gas which is in the ICM of the present-day clusters was already enriched to solar metallicity at high redshift. 
This indicates that -- as expected -- some of the cold, very metal enriched gas within the protoclusters will be heated by subsequent feedback and stays within the ICM till present-time.
Furthermore, this indicates that it could generally be possible to already detect these protoclusters at $z\approx4$ in X-ray measurements.

\section{Galaxy Cluster Evolution from protoclusters to present day}\label{sec:track}
Spectacular protocluster cores like {\it SPT2349-56} are often speculated to be the progenitors of today's most massive galaxy clusters. In the following we will use the power of the simulations and trace the previously identified protoclusters to $z=0$. 

\subsection{Cluster Mass Evolution}
\begin{figure*}
  \begin{center}
    \includegraphics[width=0.85\textwidth]{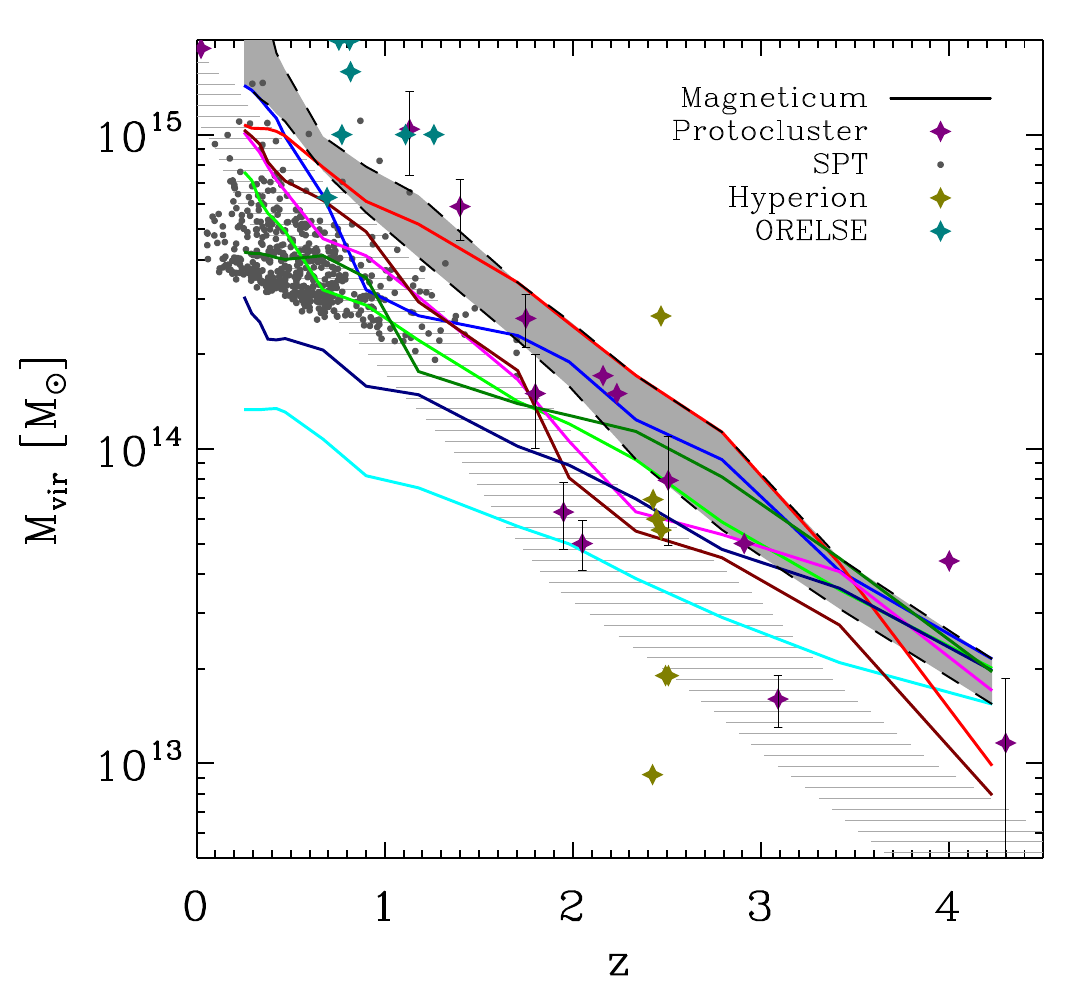}
    \caption{Total cluster mass versus redshift. The colored lines show the evolution of the protoclusters 
    selected at $z=4.2$, with colors green/\allowbreak cyan/\allowbreak blue/\allowbreak red/\allowbreak magenta/\allowbreak burgundy/\allowbreak darkblue/\allowbreak darkgreen marking the 
    eight different example-protoclusters PCl~3/12/0/5/1/7/2/4, respectively. The grey shaded area marks the 
    mass range of the 15 most massive galaxy clusters in Magneticum {\it Box~2b} at each redshift, clearly showing 
    that none of the protoclusters selected at $z=4.2$ is amongst the most massive ones at $z=0$. Grey filled 
    circles show the SPT clusters taken from \citet{miller:2018}, while the lilac stars show individual 
    protoclusters selected with X-ray or optical methods taken from \citet{miller:2018} with additional points 
    for {\it SPT-CLJ2106-5844} at $z=1.132$ from \citet{kim:2019}, the spiderweb protocluster core at $z=2.16$ from \citet{shimakawa:2014}, 
    SSA22 at $z=3.09$ from \citet{kubo:2016}, the Distant Red Core protocluster at $z=4.002$ from \citet{oteo:2018}, and RO-1001 at $z=2.91$ from \citet{daddi:2021}.
    Furthermore, the masses and redshifts of the structures from ORELSE by \citet{tomczak:2017} are shown as dark green symbols, and the values for the HYPERION group of merging protoclusters at $z\approx 2.45$ from \citet{cucciati:2018} are shown as light green symbols.
    In addition, the striped area shows the prediction from \citet{chiang:2013} from the Millenium simulation in combination with a SAM model.
    }
  {\label{fig:proto_growth}}
\end{center}
\end{figure*}

\begin{figure}
  \begin{center}
    \includegraphics[width=\columnwidth]{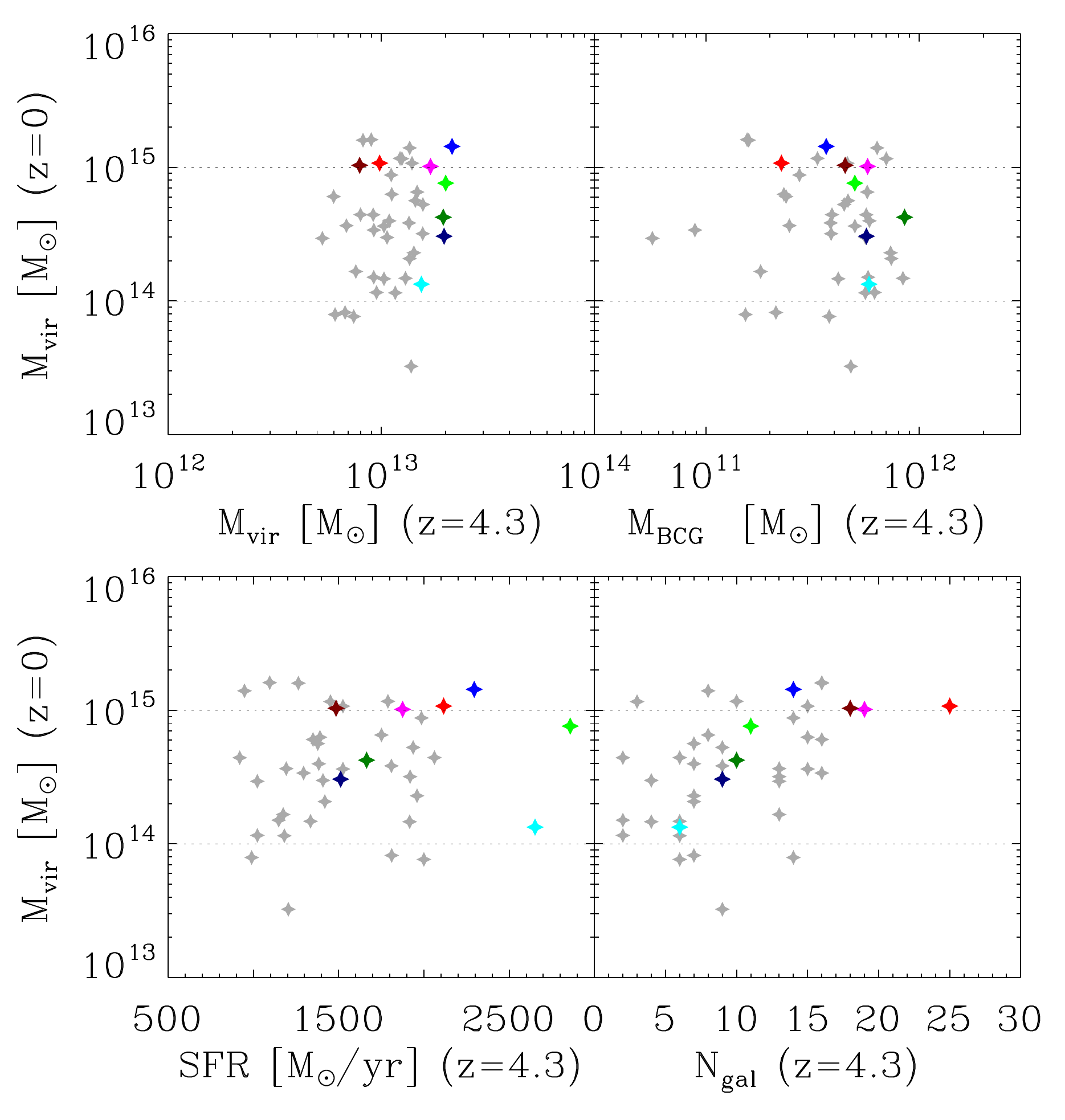}
    \caption{Virial mass at $z=0$ for the descendants of all protoclusters listed in Tab.~\ref{tab:sims1} and Tab.~\ref{tab:sims2} against the virial mass at $z=4.2$ (\textit{upper left}), the BCG stellar mass at $z=4.2$ (\textit{upper right}), the total star formation rate at $z=4.2$ (\textit{lower left}), and the richness at $z=4.2$ (\textit{lower right}). The 8 example protoclusters are marked in color, with colors green/\allowbreak cyan/\allowbreak blue/\allowbreak red/\allowbreak magenta/\allowbreak burgundy/\allowbreak darkblue/\allowbreak darkgreen marking PCl~3/12/0/5/1/7/2/4, respectively.
    }
  {\label{fig:z4z0}}
\end{center}
\end{figure}
While protoclusters surely present very large over-densities at high redshift, the complicated merging process which is involved in the formation of the most massive structures in the Universe leads to a large uncertainty for matching the most massive structures appearing at high redshift to the most massive structures observed at present time. 
To illustrate this point, Fig.~\ref{fig:proto_growth} shows the growth of structures in the simulations in comparison with observations at various redshifts. The dark gray band marks the range of virial masses of the 10 most massive systems in the simulations at the different times and very nicely encompasses the observations at various redshifts, indicated by filled symbols in that figure. 

The coloured lines show the individual evolution pathways of the 8 example protoclusters. Interestingly, none of them is among the 10 most massive systems at redshift $z=0$. Some of them even barely reach masses in the galaxy cluster range, and end up as very low mass cluster systems, like PCl~12 with a final mass of less than $2\times10^{14}M_\odot$.
Some of the most massive clusters at $z=0$ from our sample of 8 clusters are actually below a mass of $1\times10^{13}M_\odot$ at $z=4.2$, and thus below the threshold of what would be considered a protocluster (PCl~1 and PCl~7). 
From those protoclusters at $z=4.2$ that are the most massive, only PCl~0 ends up as one of the most massive clusters, although our {\it SPT2349-56} counterpart PCl~1 also reaches a final mass larger than  $1\times10^{15}M_\odot$.
\begin{figure}
  \begin{center}
    \includegraphics[width = \columnwidth]{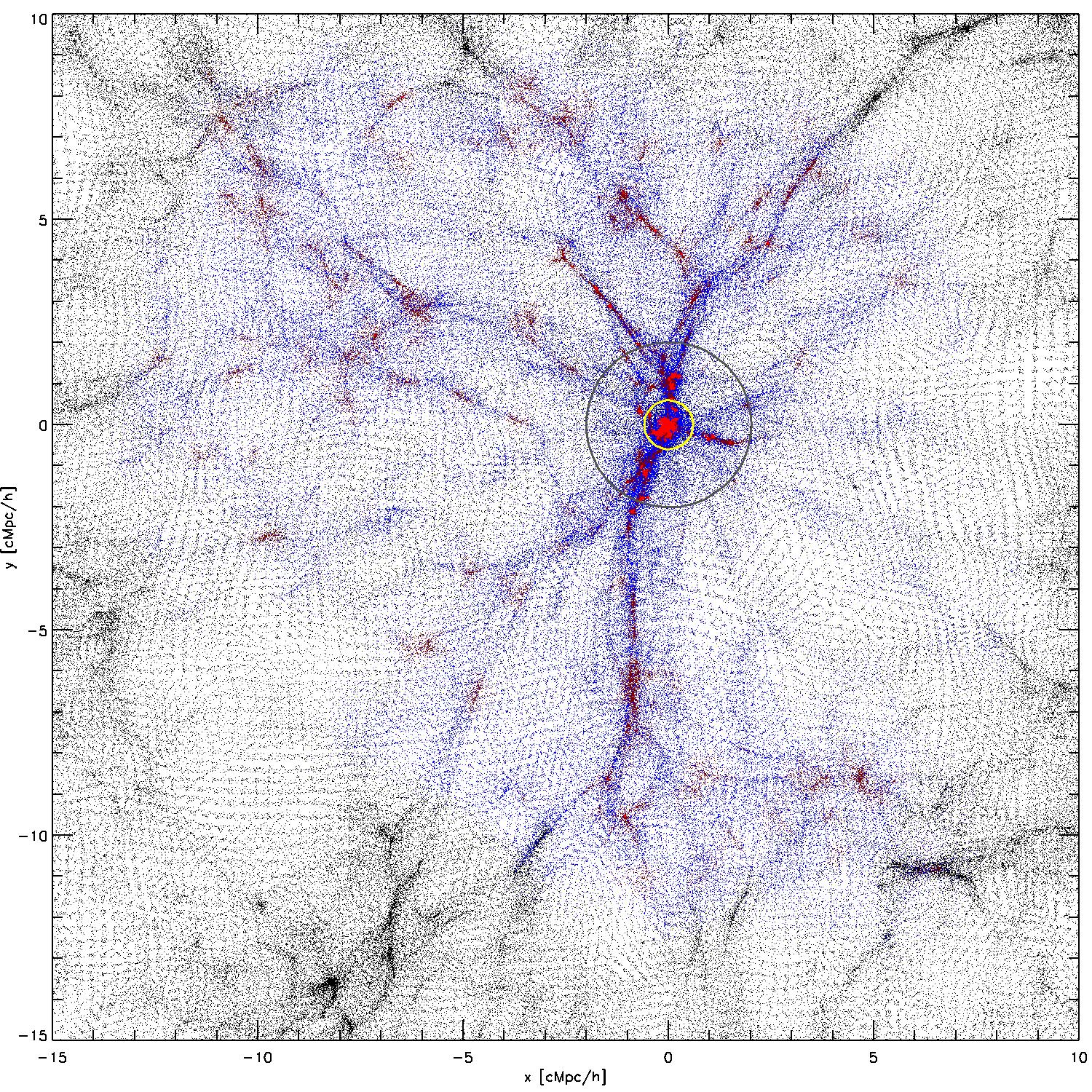}
    \caption{Slice of the Magneticum {\it Box~2b} at $z=4.2$ with a thickness of $5~\mathrm{Mpc}/h$ comoving, centered around protocluster PCl~1.
    Black dots mark all the gas particles in this slice. Everything that will end up inside the virial radius $R_\mathrm{vir}$ of the cluster at $z=0$ that evolved from this protocluster PCl~1 is highlighted in color: Gas particles are marked as blue dots, while those gas particles that end up in the cluster but are transformed into stars by $z=0$ are shown in dark red.
    Bright red dots mark the stars that already exist at $z=4.2$ and all end up in the cluster at $z=0$.
    The two circles resemble the virial radii $R_\mathrm{vir}$ at $z=4.2$ (small yellow circle) and $z=0$ (large gray circle).
    This clearly demonstrates that the protocluster volume at $z=4.2$ only provides a limited prediction for the future cluster at $z=0$.
    }
  {\label{fig:einzugsgebiet}}
\end{center}
\end{figure}

The weakness of the connection between the ranking of high-mass systems at high redshift with the ranking of the final system mass can also be clearly seen from Tab.~\ref{tab:sims1} and Tab.~\ref{tab:sims2} and the upper left panel of Fig.~\ref{fig:z4z0}.
Furthermore, Fig.~\ref{fig:z4z0} also shows that there is no correlation between the final mass of a cluster at $z=0$ and either the stellar mass of the most massive galaxy of the total star formation rate of the protocluster members. 
The only quantity for which we find a slight tendency to indicate the outcome of the mass evolution of a protocluster is the richness of the protocluster (see lower right panel of Fig.~\ref{fig:z4z0}), namely the number of galaxies that are part of the protocluster at $z=4.2$. This is further supported by what can be seen from Tab.~\ref{tab:sims2} and Tab.~\ref{tab:sims1}, namely that the only selection criterion that includes some of the most massive galaxy clusters at present day (mass rank 2 and 3 at $z=0$), is the selections by richness criterion.

An illustration of why richness is a good tracer while other quantities are not is given in Fig~\ref{fig:einzugsgebiet}.
It shows a thin ($5~\mathrm{cMpc}/h$) slice of the large scale structure around the protocluster PCl~1, for which the virial radius at $z=4.2$ is marked by the yellow circle. 
All particles of the Lagrangian volume which ends within the virial radius (large gray circle) of the descendant of PCl~1 at present day are shown in color, while the particles that do not end within the cluster are shown in gray. This demonstrates that the protocluster
at high redshifts are only the tip of the large scale structures which defines the cluster at present day. Dark red points are tracing gas
which will form stars ending in the cluster at present time, while bright red points indicate stars which have been already formed by $z=4.2$. 
As can clearly be seen, the galaxies that have already been formed are all close to the protocluster region, with only some additional stars already formed along the filaments connecting to the cluster. Thus, large numbers of member galaxies within the protocluster but also its surroundings are good tracers of the currently collapsing large scale structure, and the more such galaxies have already been formed the larger the field of influence for such a cluster.
However, the very large structure at such early times is still only partially traced by current star-formation and thus eventually not visible for observations yet. 
Nevertheless, while large richness is a good indicator for a protocluster to evolve into a massive cluster at low redshifts, a small richness is not an indicator that the protocluster cannot evolve into a massive cluster, as also visible from the lower right panel of Fig.~\ref{fig:z4z0}, as it is very possible that most of the stellar mass is still inside the collapsing filaments and not yet within the virialized area of the protocluster core.

\subsection{Forming the BCG and the ICL}
\begin{figure*}
  \begin{center}
    \includegraphics[width = .9\textwidth]{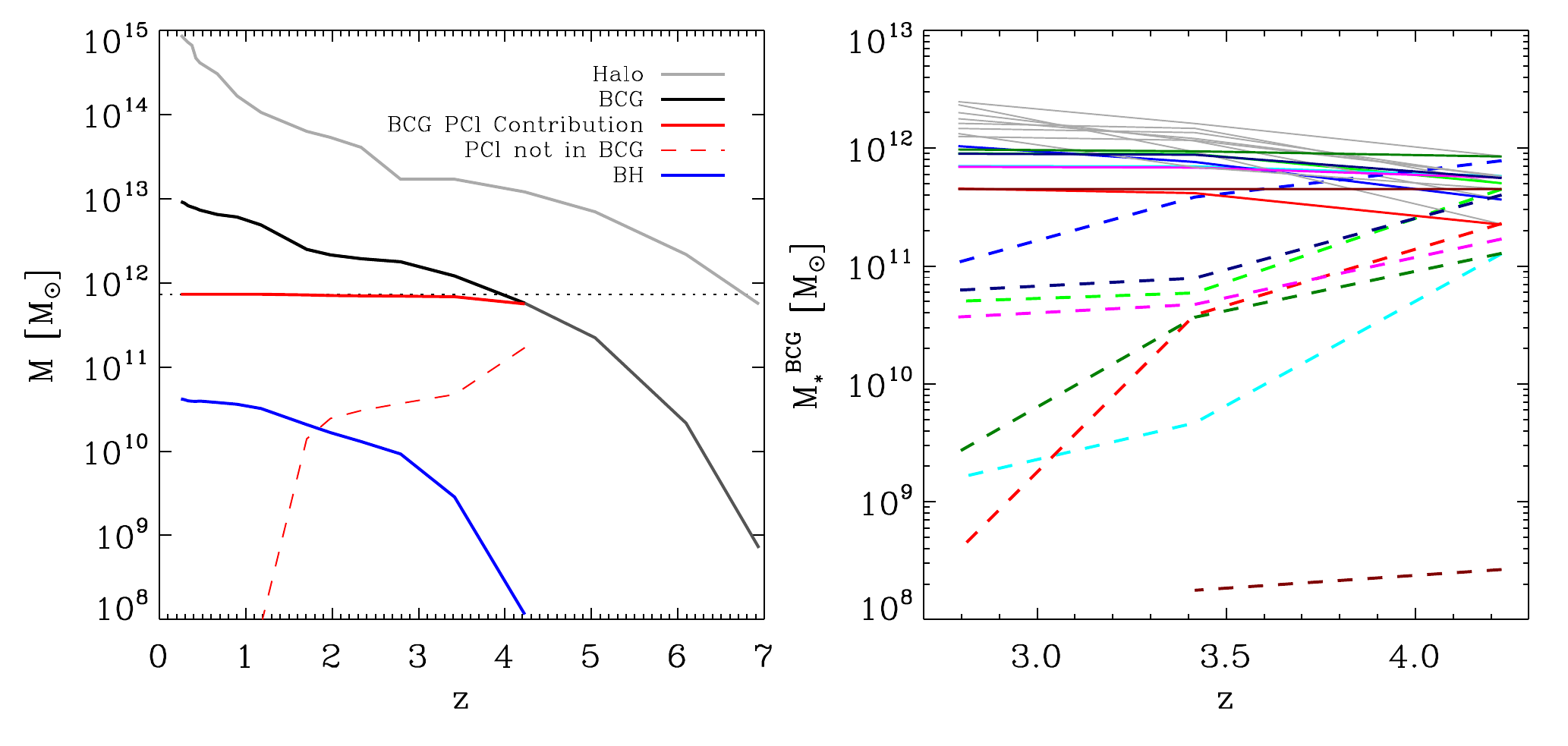}
    \caption{\textit{Left panel}: Mass growth with redshift for the example protocluster PCl~1, for the total (dark plus baryonic) mass $M_\mathrm{vir}$ (grey line), the stellar
    mass of the galaxy that is the BCG at $z=0$ (black line), and the central black hole of that galaxy (blue line).
    The solid red line shows the contribution of the protocluster stars to the BCG, with the dashed line showing the protocluster stars that are not yet inside the BCG.
    The dotted black line highlights the total stellar mass that is present already inside the protocluster PCl~1 at $z=4.2$. 
    \textit{Right panel}: Initial growth phase of the BCGs of the 8 selected example protoclusters. Gray lines show the stellar
    mass of the galaxy that is the BCG at $z=0$ (similar to the black line in the left panel), while the colored solid lines show the stellar mass already present in the protoclusters at $z=4.2$ which has been already accreted onto the BCG at the given redshift. The dashed lines show the stellar mass already present at $z=4.2$ that is not part of the BCG at the given redshifts.
    }
  {\label{fig:cl1_evol}}
\end{center}
\end{figure*}
As shown in the left panel of
Fig.~\ref{fig:cl1_evol}, the BCG of our example protocluster PCl~1 (black line) grows very rapidly by 3 orders of magnitude between $z=7$
and $z=4$, while at redshift below $z\approx1-2$ the BCG grows much slower than the halo (gray line). The stellar components of all galaxies present within the virial radius of the protocluster at $z=4.2$ are already nearly completely merged into the BCG (solid red line) at the next available snapshot at $z=3.4$, and only a very minor stellar component (dashed red line) stays either as satellites or as stripped material within the system towards lower redshift. This is true for all our 8 protocluster candidates, as shown in the right
panel of Fig.~\ref{fig:cl1_evol}, and in good agreement with the results found by \citet{rennehan:2020}. This clearly indicates a very short timescale on which these galaxies within the protocluster
will merge at this redshift. This timescale is significantly less than $0.42\mathrm{Gyr}$, which is the timespan between the stored snapshots, thus it is not possible to obtain more information on the merging processes from the simulation but the simple fact that the merging takes place. Tracing the stellar component further in time reveals that, as expected, these stars that were already part of the protoclusters at $z=4.2$ mostly end up as part of the present-day BGC, therefore effectively building up the cores of the present-day BCGs.

\subsection{Effect of cosmology and simulation volume}
\begin{figure*}
  \begin{center}
    \includegraphics[width = .45\textwidth]{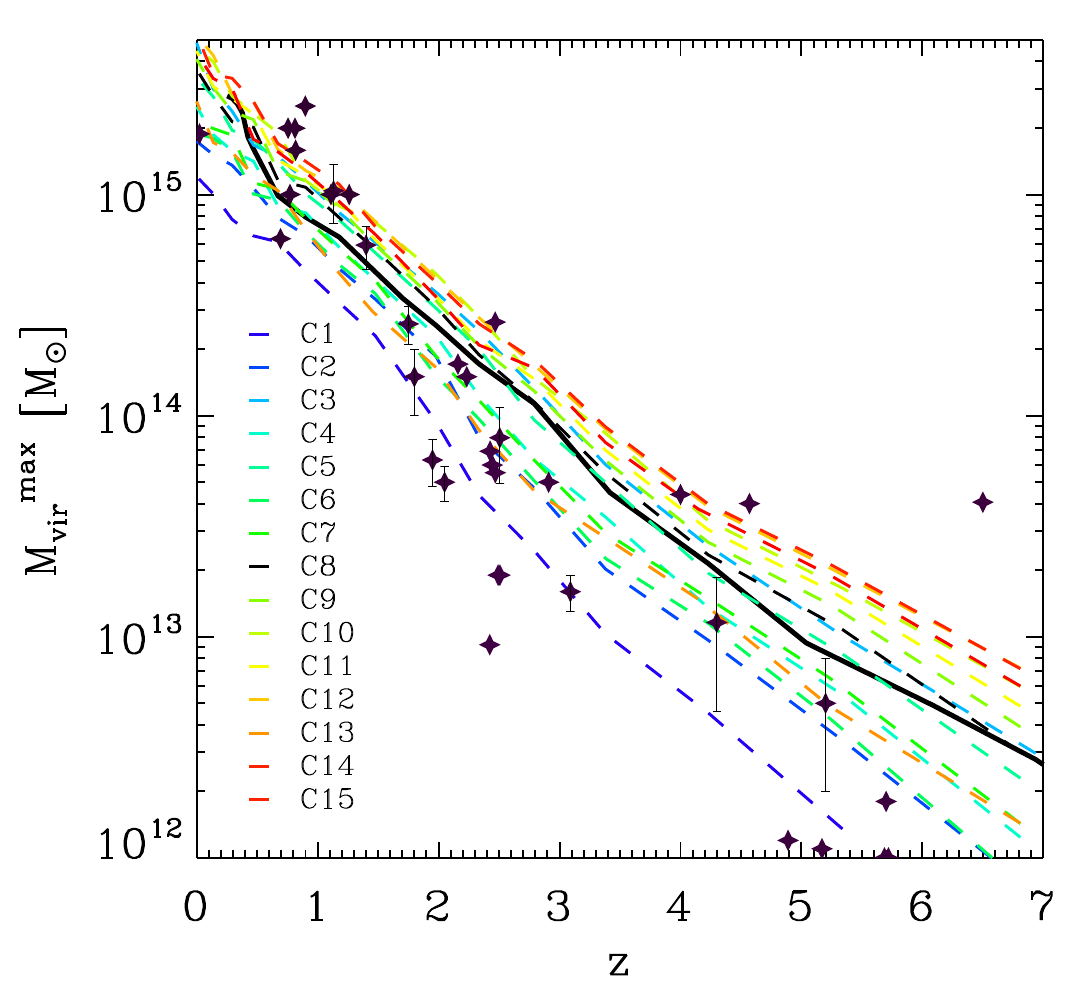}
    \includegraphics[width = .45\textwidth]{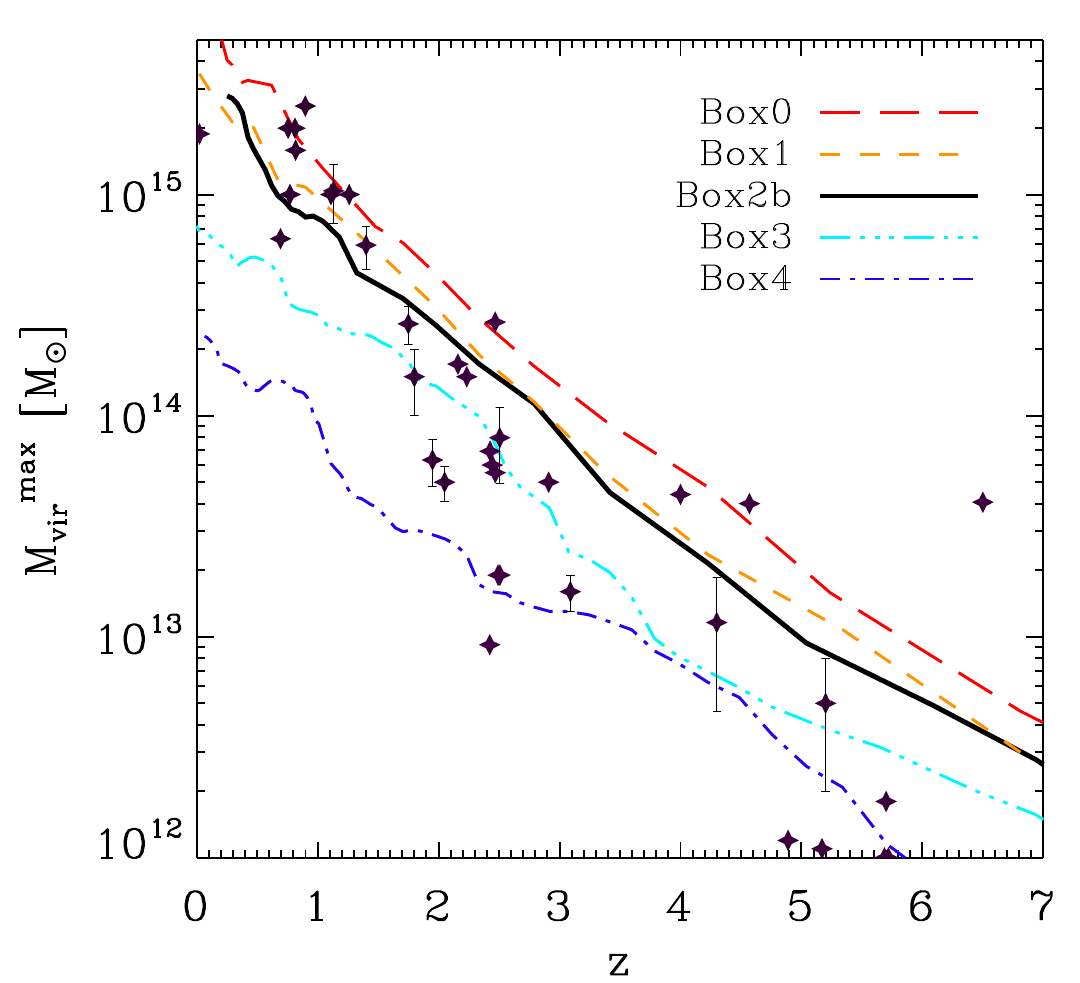}
    \caption{\textit{Left panel:} Similar to Fig.~\ref{fig:proto_growth}, but comparing the observations to the virial mass of the most massive systems from the 15 simulations with different cosmologies (see Tab.~\ref{tab:sims3}) at each redshift. Color coding according to the value of $\sigma_8$. For the detailed parameters of the models, see \citet{singh:2019}. Star symbols mark all individual (proto)cluster observations from Fig.~\ref{fig:proto_growth}, with the addition of the $z= 6.5$ protocluster from \citet{chanchaiworawit:2019} and the protocluster at $z=5.2$ from \citet{calvi:2021}. Furthermore, a compilation of lower-mass protoclusters at $z>4.8$ curtesy of F. Sinigaglia is added, sampled from observations by \citet{shimasaku:2003,venemans:2004,ouchi:2005,toshikawa:2012,toshikawa:2014,higuchi:2019,toshikawa:2020}. 
    \textit{Right panel:} Same as left panel but for the five different box volumes of the Magneticum Pathfinder simulation suite with the same cosmology.
    {\it Box~0} $(2688~\mathrm{Mpc}/h)^{3}$ and {\it Box~1} $(896~\mathrm{Mpc}/h)^{3}$ are performed with the lowest resolution, {\it Box~2b} $(640~\mathrm{Mpc}/h)^{3}$ and {\it Box~3} $(128~\mathrm{Mpc}/h)^{3}$ are performed on high resolution, while {\it Box~4} $(48~\mathrm{Mpc}/h)^{3}$ is on the highest available resolution.
    }
  {\label{fig:mass_growth_cosmo}}
\end{center}
\end{figure*}
\begin{table}
\caption{Cosmological parameters of the 15 different cosmological runs of {\it Box1~mr}, as presented by \citet{singh:2019}.
The gray shaded row corresponds to the setting used for the main suit of the Magneticum simulations, especially {\it Box~2b hr}
which is used for the major part of this study.
}
\label{tab:sims3}
\def\colorrow{\noalign{\color[gray]{0.9}\hrule height 13pt}\\[-26pt]}
\begin{center}
\begin{tabular}{c|ccccc}
\hline\hline
   & $\Omega_{0}$ & $\Omega_\mathrm{b}$ & $\sigma_\mathrm{8}$ & $\mathrm{H}_\mathrm{0}$ & $\mathrm{f}_\mathrm{b}$ \\
\hline
C1  & 0.153 & 0.0408 & 0.614 & 66.6 & 0.267 \\
C2  & 0.189 & 0.0455 & 0.697 & 70.3 & 0.241 \\
C3  & 0.200 & 0.0415 & 0.850 & 73.0 & 0.208 \\
C4  & 0.204 & 0.0437 & 0.739 & 68.9 & 0.214 \\
C5  & 0.222 & 0.0421 & 0.793 & 67.6 & 0.190 \\
C6  & 0.232 & 0.413  & 0.687 & 67.0 & 0.178 \\
C7  & 0.268 & 0.0449 & 0.721 & 69.9 & 0.168 \\
\colorrow
C8  & 0.272 & 0.0456 & 0.809 & 70.4 & 0.168 \\
C9  & 0.301 & 0.0460 & 0.824 & 70.7 & 0.153 \\
C10 & 0.304 & 0.0504 & 0.886 & 74.0 & 0.166 \\
C11 & 0.342 & 0.0462 & 0.834 & 70.8 & 0.135 \\
C12 & 0.363 & 0.0490 & 0.884 & 72.9 & 0.135 \\
C13 & 0.400 & 0.0485 & 0.650 & 67.5 & 0.121 \\
C14 & 0.406 & 0.0466 & 0.867 & 71.2 & 0.115 \\
C15 & 0.428 & 0.0492 & 0.830 & 73.2 & 0.115 \\ 
\hline
\end{tabular}
\end{center}
\end{table}
Finally, we evaluate the expected mass of the most massive structures at different redshifts, depending on the underlying cosmology. Therefore, we use a new extension of the {\it Magneticum} simulation set as presented by \citet{singh:2019}. These are re-simulations of {\it Box~1a}, which has a volume of $(896 h^{-1}\mathrm{Mpc})^3$, with $2\times1526^3$ particles, using 15 different cosmologies by varying
$\sigma_8$, $\Omega_0$, $H_0$ as well as $\Omega_b$. The cosmological parameters for the different runs are listed in Tab.~\ref{tab:sims3}. All simulations are run on the lowest available resolution (mr), with a particle resolution of $m_\mathrm{DM} = 1.3\times10^{10} M_{\odot}/h$ and $m_\mathrm{Gas} = 2.9\times10^{9} M_{\odot}/h$ for dark matter and gas, respectively. For details see \citet{singh:2019}.

Although these simulations follow the same hydrodynamical treatment and sub-grid physics description as the simulation used for the previous part of this study, they have significantly less resolution and therefore we cannot perform a detailed analysis of protoclusters as done before with the much higher resolution simulation of {\it Box~2b}. 
Nevertheless, the resolution is high enough to predict the expectation of the most massive viral mass as function redshift, similar to the upper limit of the gray shaded area in Fig~\ref{fig:proto_growth}. 
The colors are the same as used by \citet{singh:2019}, and were originally chosen to go from low $\Omega_m$ (blue) to a high value of $\Omega_m$ (red). 
Due to choosing the cosmological parameters to follow the current observational constrains within the different figures of merit (for details see \citet{singh:2019}), this means that the different values have non trivial relationships and therefore it is impossible to construct a strict hierarchy of the models. 
As can be seen from left panel of Fig.~\ref{fig:mass_growth_cosmo}, the absolute mass of the most massive clusters at high redshift depends mostly on the value of $\sigma_8$, i.e., the larger $\sigma_8$ the more massive structures appear already at high redshift, as can be seen especially from the runs C~3 (turquoise line) and C~13 (orange line).
Effectively, this results in a different slope for the highest-mass per redshift relation, with flatter slopes for larger values of $\sigma_8$.
There is only a small trend with $\Omega_m$, that is larger $\Omega_m$ producing larger structures at a given redshift, but this trend is only of secondary order.
This clearly shows that measuring the most massive structures present at different redshift can set constraints on the values of $\sigma_8$, and thus protoclusters might be useable as cosmological probes.

Furthermore, we can also see the effect of the different volumes of the simulated boxes from the left panel of Fig.~\ref{fig:mass_growth_cosmo}, where the solid black line marks the most massive halos per redshift for the smaller but higher resolved simulation {\it Box~2b}, in comparison to the black dashed line that marks the {\it Box~1a} cosmological run (C~8) with the same cosmology as the {\it Box~2b} simulation used for most of this work.
The low resolution simulation of {\it Box~1a} (dashed line) has roughly 3 times the volume of the high resolution simulation {\it Box~2b} (solid line), showing a mild trend to harbour larger structures,
as expected due to the larger cosmological modes included in the larger simulation volumes, independent of the resolution. 
This is also further highlighted in the right panel of Fig.~\ref{fig:mass_growth_cosmo}, where the most massive structures per redshift are shown for all five volumes that are part of the Magneticum pathfinder simulation suite (with the same cosmology as {\it Box~2b}).
This clearly demonstrates the need for large enough box volume simulations to capture those massive protoclusters that are now detected at high redshifts and study their properties and evolution pathways.

\subsection{Predictions to $z=10$}
\begin{figure}
  \begin{center}
    \includegraphics[width = .45\textwidth]{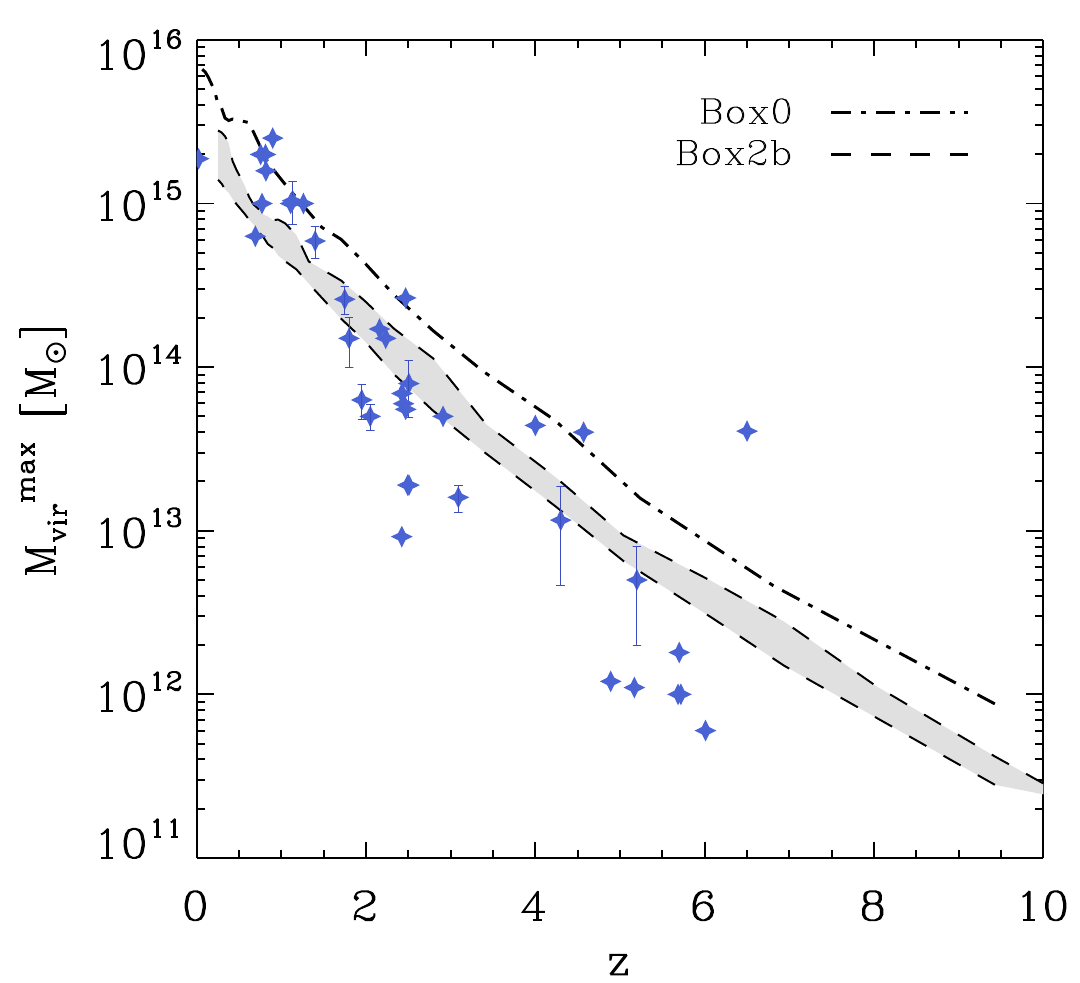}
    \caption{Similar to Fig~\ref{fig:proto_growth}, but extended as predictions of the highest expected total masses to be found up to redshifts of $z=10$, using the two largest box volumes from the Magneticum pathfinder suite of simulations at low {\it Box~0} $(2688~\mathrm{Mpc}/h)^{3}$ and high {\it Box~2b} $(640~\mathrm{Mpc}/h)^{3}$ resolutions.
    For {\it Box~2b}, at each redshift the mass range of the 10 most massive systems is shown as gray shaded area, while for {\it Box~0} only the line for the most massive system per redshift is shown.
    A compilation of observed systems as described in Fig~\ref{fig:proto_growth} with the addition of the high redshift $z= 6.5$ protocluster from \citet{chanchaiworawit:2019} and the protocluster at $z=5.2$ from \citet{calvi:2021} is shown as blue stars.
    Furthermore, a compilation of lower-mass protoclusters at $z>4.8$ curtesy of F. Sinigaglia is added, sampled from observations by \citet{shimasaku:2003,venemans:2004,ouchi:2005,toshikawa:2012,toshikawa:2014,higuchi:2019,toshikawa:2020}.
    }
  {\label{fig:mass_growth_resol}}
\end{center}
\end{figure}
Finally, we extend our study of the most massive already bound systems with redshift towards even higher redshifts of $z=10$, the highest redshift at which already bound halos with stellar components can be found in the simulations given the resolution limits. 
We predict that halo masses of up to $M_\mathrm{vir} = 1\times10^{12}M_\odot$ can already be found at $z=9$, as shown in Fig.~\ref{fig:mass_growth_resol}, and a few times $10^{11}M_\odot$ at $z=10$. 
As can also be seen, the massive protocluster at $z= 6.5$ presented by \citet{chanchaiworawit:2019} has a mass larger than any or our simulations, even the largest simulation volume, can reproduce, albeit the error bars are rather large. 
Whether this indicates that none of our simulation volumes is large enough to capture such massive structures, if that particular object is an outlier, or if the mass could be overestimated is beyond the scope of this work, but will be interesting to investigate in the future.
All other systems reported at redshifts above $z=4.2$ are well within the predicted mass range from our simulation.

\section{Discussion and Conclusion}\label{sec:conclusion}
Utilizing the very large cosmological hydrodynamical simulation {\it Box~2b} of the {\it Magneticum} project, which includes a sub-resolution treatment of star formation, stellar evolution and a treatment of the effect of super massive black holes, we show that such simulations, once the volume is large enough, can successfully produce massive protoclusters like {\it SPT2349-56}, reproducing many of their physical properties. 
Especially, we find several virialized structures at $z=4.2$ with a similar number of member galaxies (richness) and the same dynamical properties of the member galaxies as the observed protocluster {\it SPT2349-56} at $z=4.3$ \citep{miller:2018,rotermund:2021}, but also similar to the protocluster core found by \citet{oteo:2018} at $z=4$. 
The simulations also predict that several of the member galaxies of these structures are fast rotating systems, in agreement with the observed findings for member galaxies of {\it SPT2349-56}.

We find that these member galaxies of the protoclusters at $z=4.2$ will merge on very short timescales, forming the progenitors of today's cluster BCGs, in agreement with results found by \citet{rennehan:2020}. 
The stellar component merging from these satellite galaxies assembles to build up the core of the BCG, and indicating that the centres of today's massive BCGs echo this fast growing phase and their central velocity dispersion may reflect the dynamics of the galaxies within these protoclusters, being significant smaller than the velocity dispersion of the present-day clusters the BCG lives in \citep[see][]{sohn:2019,bender:2015,remus:2017}.

However, while the general stellar and total masses of the simulated protoclusters and the number and dynamics of member galaxies within resemble the observations closely, the instantaneous integrated star-formation rate within such simulated protoclusters is a factor $\approx2\dots 3$ smaller than the observed values, although the simulations are reproducing the observed integrated gas mass within the protoclusters.
As the simulations also reasonably well reproduce the main sequence of star-forming galaxies as well as the observed stellar mass function up to this redshift, this indicates that the star-formation in the simulations lacks the ability to reproduce higher star-formation efficiency (or accordingly lower depletion timescales) at least in certain environments.
The reason for this issue can be found most likely in the implementation of the star formation process that currently assumes a rather continuous star formation and a Schmitt-Kennicutt relation that holds true even at high redshifts, an issue that needs to be addressed in future simulations.

While star forming galaxies are detectable in the high redshift protoclusters due to their large gas reservoirs, the quiescent galaxies at high redshifts are much more elusive.
Using our sample of 42 simulated protoclusters, we find that there can already be quenched galaxies in such protoclusters, albeit their number is low, and on average at the total mass range of such protoclusters (around $10^{13}M_\odot$) barely at $10\%$. 
Their quiescent fractions do not depend on the dynamical state of the protocluster, neither its virial mass nor the overall star formation rate or the richness.
We compare our quiescent fractions to observations at lower redshifts where quiescent fractions can actually be inferred \citep{strazzullo:2019,sarron:2021}, and find an overall agreement, in agreement with previous work done by \citet{lotz:2019,lotz:2021}.
We find the quiescent fraction to overall strongly increase with decreasing redshift, even at fixed total mass, clearly showing that quenching becomes more efficient at lower redshifts in both group and cluster environments.

The simulations predict that the star-forming gas in protoclusters at redshifts of $z \approx 4$ is already enriched to roughly solar values. 
Part of this cold gas is expected to be subsequently heated by feedback and become part of the intra-cluster medium of the forming galaxy cluster. 
About half of the gas within these structures at $z=4.2$ is already significantly heated to temperatures around $1\mathrm{keV}$, and a very small fraction ($\approx 2$\%) of this hot gas is already enriched to one tenth of the solar value.

Using the full power of the simulation, we traced the protoclusters identified at $z=4.2$ down to $z=0$, to test the hypothesis that the extremely massive structures found at high redshifts really are the progenitors of the most massive galaxy clusters at present day.
However, at $z\approx4$, these protocluster regions reflect only a very minor part of the Lagrangian region which will collapse into the final galaxy clusters at $z=0$, and therefore none of the protocluster properties at $z=4.2$ (e.g., virial mass, star-formation rate, stellar mass, or richness in members) proves to be a good proxy for the mass of the final cluster at $z=0$. 
In fact, from our 8 examples chosen to be a among the top in these measures at $z=4.2$, none is among the 10 most massive clusters at $z=0$. 
Even more striking, one of them evolves barely into a very low mass cluster with a virial mass of less than $1\times10^{14}M_\odot$.
From the full sample of 42 protoclusters at $z=4.2$, four do not even grow above the mass of $1\times10^{14}M_\odot$.
This is due to the fact that nodes in the cosmic web can collapse at very different timescales, and some of the nodes that collapse rather early starve and become fossil systems.
With respect to the observable quantities within protoclusters at high redshift, we find the richness in galaxy members to be the only quantity that has a slight indication for the future of the system, in that 
rich systems tend to also evolve into massive systems at $z=0$, however, the opposite is not true as also some systems that are low in richness at $z=4.2$ evolve into massive clusters at present-day.

Utilizing a second set of simulations from the Magneticum pathfinder suite of simulations, which were performed with 15 different cosmologies \citep{singh:2019}, we quantified the expected largest mass of bound systems with redshift.
Given the current uncertainties in the actual values of the cosmological parameters, we showed that the highest mass per redshift is actually a good tracer for $\sigma_8$, with the discrepancies larger the higher the redshift.
Thus, finding the most massive bound systems at high redshifts can set constraints on those cosmological parameters.

To conclude, we found that, on one hand, various aspects of observed protoclusters can be successfully reproduced by current state-of-the-art cosmological simulations if the simulation volumes are large enough. 
These protocluster structures have already virialized cores, already hosting a significant hot atmosphere which could be targeted observationally. 
On the other hand, detailed comparisons of the star-formation rates reveal, as indicated in previous work from various simulation suites, that there seems to be some environmental increase in star-formation efficiency (or accordingly reduced depletion timescales) which current sub-resolution models describing the star-formation within the simulations are not able to capture, albeit gas masses and both the general star formation main sequence and the stellar mass functions are reproduced successfully.
While some of the protocluster systems found at $z\approx4$ evolve into massive clusters at $z=0$, they are not among the progenitors of today's most massive clusters, and some of them barely reach cluster masses.
Overall, the simulations shape a picture of a fast growing mode of massive systems at early time, which should still be echoed in the dynamical properties of the central parts of today's massive BCGs, with a broad range of outcome in total mass at low redshifts.

\section*{Acknowledgements}
We especially thank Scott Chapman for several helpful discussion, Florian Sarron for providing the data from the Detectivz survey in mass bins comparable to the simulated cuts, and Olivier Gilbert for providing the COSMOS data in redshift bins comparable to the simulation outputs.
Furthermore, we would like to thank Francesco Sinigaglia for compiling and providing a high redshift protocluster sample, and Giulia Rodighiero for helpful comments.
The {\it Magneticum} simulations were performed at the Leibniz-Rechenzentrum with CPU time assigned to the Project {\it pr83li}. This work was supported by the Deutsche Forschungsgemeinschaft (DFG, German Research Foundation) under Germany's Excellence Strategy -- EXC-2094 -- 390783311. KD acknowledges support by the COMPLEX project from the European Research Council (ERC) under the European Union's Horizon 2020 research and innovation program grant agreement ERC-2019-AdG 882679. 
HD acknowledges financial support from the Agencia Estatal de Investigaci\'on del Ministerio de Ciencia e Innovaci\'on (AEI-MCINN) under grant \textit{La evoluci\'on de los c\'iumulos de galaxias desde el amanecer hasta el mediod\'ia c\'osmico} with reference PID2019-105776GB-I00/DOI:10.13039/501100011033, and support from the
ACIISI, Consejer\'ia de Econom\'ia, Conocimiento y Empleo del Gobierno de Canarias and the European Regional Development Fund (ERDF) under grant with reference PROID2020010107.
We are especially grateful for the support by M. Petkova through the Computational Center for Particle and Astrophysics (C2PAP).

\vspace{2\baselineskip}
\bibliography{bibliography}
\bibliographystyle{aa}

\label{lastpage}
\end{document}